\documentclass[conference]{IEEEtran}
\IEEEoverridecommandlockouts

\usepackage{cite}
\usepackage{amsmath,amssymb,amsfonts}
\usepackage{graphicx}
\usepackage{textcomp}
\usepackage{xcolor}
\usepackage{url}
\usepackage[hidelinks]{hyperref}
\hypersetup{
  pdftitle={Calibrated Decision Models for Autonomous Penetration-Testing Harnesses: JEV and Laya as System One Decision Layers for LLM-Driven Pentest Agents},
  pdfauthor={Joas Antonio dos Santos},
  pdfsubject={System One decision models; autonomous penetration testing; calibrated severity; JEV; Laya; NeuroSploit},
  pdfkeywords={autonomous penetration testing, System One, JEV, Laya, calibrated decisions, RLHF, RLAIF, RLCD, offensive security, LLM agents}
}
\usepackage{booktabs}
\usepackage{multirow}
\usepackage{listings}
\usepackage{tikz}
\usetikzlibrary{shapes.geometric, arrows.meta, positioning, fit, backgrounds, calc}
\usepackage{float}
\usepackage{enumitem}
\usepackage{placeins}

\lstdefinelanguage{Rust}{
  morekeywords={fn, let, mut, pub, struct, impl, use, async, await, match, if, else, for, in, return, self, crate, mod, type, where, trait, enum, const, static, ref, move, loop, while, break, continue, as, true, false, Some, None, Ok, Err},
  sensitive=true,
  morecomment=[l]{//},
  morecomment=[s]{/*}{*/},
  morestring=[b]",
}

\newcommand{\cmark}{\checkmark}

\begin{document}

\title{Calibrated Decision Models for Autonomous\\Penetration-Testing Harnesses:\\JEV and Laya as System One Decision Layers\\for LLM-Driven Pentest Agents}

\author{\IEEEauthorblockN{Joas Antonio dos Santos}
\IEEEauthorblockA{\textit{Independent Researcher --- AI and Offensive Security} \\
S\~ao Paulo, SP, Brazil \\
joas.santos@redteamleaders.com \\
ORCID: 0009-0003-1772-9826}
}

\maketitle

\begin{abstract}
Autonomous penetration-testing harnesses rely on large language models (LLMs) for reconnaissance, exploit generation, and report writing, yet delegate critical adjudication decisions---finding confirmation, severity grading, agent selection---to the same generative models that produce them. This coupling introduces calibration failures: inflated severity scores driven by vulnerability class rather than demonstrated impact, false-positive findings accepted on narrative plausibility, and wasted compute on irrelevant attack surfaces. We investigate the integration of \emph{System One} decision models---lightweight, non-generative classifiers that return typed, probabilistically calibrated verdicts---into the decision layer of an autonomous pentest harness. We present five contributions. First, we formalize four decision points where a System One model replaces free-form LLM judgment with structured, auditable verdicts: finding adjudication, severity recalibration, agent pruning, and confirmation loops. Second, we report an exploratory case study on NeuroSploit, an open-source Rust harness, comparing single runs with and without TypeSafe System One (Jev) against a 13-vulnerability web target; the observations---differences in severity distribution, wall-clock time, and data-type-aware grading---motivate the architecture but do not constitute a controlled experiment with statistical power. Third, we survey the emerging landscape of System One models---the proprietary Jev family (including the browser-optimized Jev-Ultrafast) and the open-source Laya---and report their published specifications without extrapolating cross-benchmark comparisons to the pentest domain. Fourth, we analyze the training paradigms underlying calibrated decision models (RLHF, RLAIF, RLCD, RLHV) and their implications for trust in security-critical pipelines. Fifth, as future work we sketch \emph{Rave}: a domain-adapted System One variant to be fine-tuned on offensive-security decision distributions, with proposed training data requirements, evaluation protocol, and projected impact on harness assurance properties.
\end{abstract}


\section{Introduction}\label{sec:intro}

The capability of LLM-driven penetration-testing agents has advanced rapidly. Multi-agent harnesses such as MAPTA~\cite{david2025mapta} achieve 76.9\% success on the 104-challenge XBOW benchmark, while plain coding agents with frontier models reach 92.3\% under model scaling alone~\cite{dhakal2026baselines}. Mayoral-Vilches proposes a six-level autonomy taxonomy adapted from SAE~J3016, placing current systems at Levels~3--4 and noting that even headline systems require human review before vulnerability submission~\cite{mayoral2025gap}. Curtis and Eisty's systematic review of 58 studies finds reinforcement learning dominant at 77\% of reviewed works, with LLM-based approaches emerging but real-world deployment sparse~\cite{curtis2025role}. Nguyen and Husain demonstrate that agentic AI systems exhibit a 58.5\% attack success rate across 130 test cases, with framework choice significantly affecting refusal rates~\cite{nguyen2025agenticpentest}.

These advances are measured almost exclusively by \emph{capability}---whether the agent captures a flag or produces a proof-of-concept. A complementary dimension, \emph{assurance}, concerns whether the output is true, whether the agent stayed within scope, and whether the engagement can withstand post-hoc scrutiny~\cite{santos2026assurance}. Santos defines five assurance properties (evidence grounding, non-destructive claim reduction, computed severity, enforced authorization, and tamper-evident accountability) and argues they belong to the harness, not the model.

This paper addresses a specific failure mode within that assurance gap: the \emph{decision bottleneck}. At four critical junctures---finding adjudication, severity computation, agent selection, and confirmation loops---the harness must convert evidence into a typed verdict. When this verdict is produced by the same generative LLM that wrote the exploit narrative, three pathologies emerge:

\begin{enumerate}[leftmargin=*]
\item \textbf{Calibration collapse.} The model's confidence in its own output does not track the probability that the claim is true; a fluent narrative about a Critical SQL injection may rest on a reflected parameter that was escaped.
\item \textbf{Class-driven severity.} Severity is assigned by vulnerability class (``SQL injection is Critical'') rather than by demonstrated impact, inflating scores and eroding client trust.
\item \textbf{Compute waste.} Agent selection driven by LLM reasoning spends tokens evaluating surfaces that a lightweight classifier could dismiss in milliseconds.
\end{enumerate}

We propose replacing the generative LLM at these decision points with a \emph{System One} model: a non-autoregressive classifier that evaluates typed questions against a state and returns structured, probabilistically calibrated verdicts. The term ``System One'' borrows from Kahneman's dual-process theory~\cite{kahneman2011thinking}---fast, bounded, pattern-matching judgment as opposed to slow, deliberative reasoning---and has been adopted by TypeSafe AI for their Jev model family and by the open-source Laya project.

The remainder of this paper is organized as follows. Section~\ref{sec:background} covers the harness landscape and reinforcement-learning paradigms. Section~\ref{sec:sysone_arch} presents System One primitives and their offensive-security semantics. Section~\ref{sec:neurosploit} describes the NeuroSploit harness architecture. Section~\ref{sec:integration} formalizes the four decision points. Section~\ref{sec:speed} analyzes latency and speed implications. Section~\ref{sec:calibration} examines probability calibration in security contexts. Section~\ref{sec:benchmark} reports the exploratory case study. Section~\ref{sec:scenarios} details offensive-security application scenarios. Section~\ref{sec:landscape} surveys the System One model landscape. Section~\ref{sec:proposal} sketches Rave as future work. Sections~\ref{sec:discussion}--\ref{sec:conclusion} discuss limitations, related work, and conclusions. Table~\ref{tab:status} clarifies the status of each contribution.

\begin{table}[h]
\centering
\caption{Contribution status: what has been implemented, measured, or proposed.}
\label{tab:status}
\small
\resizebox{\columnwidth}{!}{%
\begin{tabular}{@{}llp{3.5cm}@{}}
\toprule
\textbf{Contribution} & \textbf{Status} & \textbf{Evidence} \\
\midrule
DP1--DP4 formalization & Implemented & NeuroSploit harness code \\
Jev integration & Implemented & \texttt{--typesafe} flag \\
Case study (single run) & Measured & 1 run/condition, 1 target \\
Re-test (post-fix harness) & Measured & Separate harness version \\
Jev vs Laya comparison & Reported & Published specs (not paired) \\
Decision-theoretic framework & Proposed & Formal analysis \\
RLHV training paradigm & Proposed & Design only \\
Rave (domain-adapted model) & Future work & Design only \\
\bottomrule
\end{tabular}%
}
\end{table}

\section{Background}\label{sec:background}

\subsection{Autonomous penetration-testing harnesses}

An autonomous pentest harness is the runtime wrapping an LLM agent with tool access, scope enforcement, memory, and output parsing. The harness---not the model---determines whether a reported finding is grounded in evidence, whether scope was enforced in code, and whether an audit trail exists. Santos~\cite{santos2026assurance} formalizes five assurance properties and shows that capability and assurance are orthogonal: the same base model in two harnesses can differ by tens of percentage points on the same benchmark~\cite{cybench2024,wang2025checkmate}, establishing the harness as the decisive engineering surface.

Representative systems span single-agent ReAct loops~\cite{fang2024oneday}, multi-agent planners (CHECKMATE~\cite{wang2025checkmate}), tool-grounded executors (MAPTA~\cite{david2025mapta}), and reproducible trace harnesses. Dhakal et al.~\cite{dhakal2026baselines} show that purpose-built architectures add 5--10 percentage points over plain coding agents when models are matched, but model scaling narrows this gap substantially.

The fundamental tension in harness design is between recall (finding vulnerabilities) and precision (reporting only real ones). A harness that maximizes recall by accepting every LLM-generated narrative produces reports that clients cannot trust; one that maximizes precision by requiring deterministic proof misses real vulnerabilities whose evidence is ambiguous. System One models offer a principled middle ground: calibrated probability estimates that allow the harness to set explicit thresholds based on the engagement's risk tolerance.

\subsection{Reinforcement learning paradigms for calibrated models}\label{sec:rl_background}

The training of calibrated decision models draws on several reinforcement learning paradigms, each with distinct implications for trust in security-critical applications.

\textbf{RLHF} (Reinforcement Learning from Human Feedback)~\cite{ouyang2022training} trains a reward model from pairwise human preferences, then optimizes a policy against that reward. RLHF has become the standard alignment technique for frontier LLMs but introduces well-documented pathologies: reward hacking, where the policy exploits distributional gaps in the reward model; preference noise, where annotator disagreement propagates to the reward signal; and mode collapse, where optimization narrows output diversity~\cite{casper2023open}. In a severity-grading context, RLHF-trained models learn that annotators prefer confident-sounding severity labels, creating systematic overconfidence that inflates vulnerability scores.

\textbf{RLAIF} (Reinforcement Learning from AI Feedback)~\cite{bai2022constitutional} replaces human annotators with an AI model that generates preference labels. RLAIF reduces annotation cost and scales to larger datasets, but the calibration of the labels inherits the calibration failures of the labeling model. If the labeling model is itself RLHF-trained with overconfidence bias, the trained model inherits and potentially amplifies that bias. This circular dependency makes RLAIF unsuitable as a sole training signal for security-critical decisions where miscalibrated confidence can trigger incorrect remediation priorities.

\textbf{RLCD} (Reinforcement Learning for Calibrated Decisions)~\cite{typesafe2026docs,laya2026} trains against strictly proper scoring rules~\cite{gneiting2007strictly}---loss functions where the unique optimal prediction is the true probability distribution. The Brier score, $\mathrm{BS} = \frac{1}{N}\sum_{i=1}^{N}(p_i - o_i)^2$ (where $p_i$ is the predicted probability and $o_i \in \{0,1\}$ is the outcome), is the canonical example. A model minimizing the Brier score is incentivized to report $p = 0.7$ when the event occurs 70\% of the time, not to round to 1.0 because ``confident'' is preferred. TypeSafe describes Jev as trained with an undisclosed RLCD methodology~\cite{typesafe2026docs}; Laya documents training against the Brier score explicitly~\cite{laya2026}.

The distinction between RLHF and RLCD is not academic for offensive security. Consider a finding where the evidence shows that a SQL injection payload reached the interpreter but no data was extracted. An RLHF-trained model, having learned that ``SQL injection'' is associated with ``Critical'' in human preferences, may rate this Critical with high confidence. An RLCD-trained model, having learned that the conditional probability of data extraction given interpreter access is approximately 0.6 in its training distribution, reports that probability, enabling the harness to grade severity on the demonstrated impact (reached) rather than the assumed impact (extracted).

\textbf{RLHV} (Reinforcement Learning from Human Verification) extends RLCD for harness-integrated models. The verification signal comes not from preferences or AI feedback but from \emph{deterministic outcome verification}---in the pentest context, from per-CWE validators that confirm or reject findings based on evidence. The training loop couples the decision model to the harness's own grounding machinery, creating a closed feedback cycle where the validator's verdict is the ground truth for the scoring rule.

The RLHV paradigm has a deeper implication for autonomous pentest systems: it creates a \emph{self-improving} harness. Each engagement produces (prediction, validator-verdict) pairs that serve as training data for the next model iteration. Over time, the decision model converges toward the validators' decision boundaries, making its predictions increasingly aligned with the harness's own grounding criteria. This convergence is guaranteed by the proper scoring rule: the Brier score's unique minimum is the true conditional probability, so the training objective directly aligns the model's output with the validator's behavior distribution.

The circularity concern---the model learns to predict the validator, not the ground truth---is addressed by the validators' design. NeuroSploit's 27 CWE validators are deterministic functions of evidence: they check for specific patterns (SQL error strings, reflected payload markers, authorization token reuse) that constitute proof of exploitability. The validator is correct by construction for supported evidence types; the model learns to predict this correctness, which is the desired behavior.

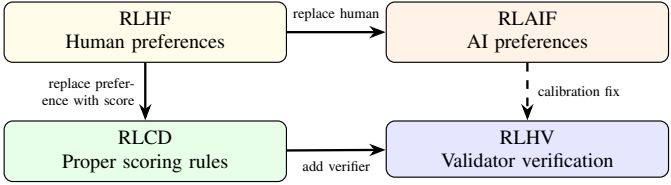
\begin{figure}[t]
\centering
\resizebox{\columnwidth}{!}{%
\begin{tikzpicture}[
    node distance=0.7cm,
    box/.style={rectangle, draw, rounded corners=3pt, text width=3.2cm, minimum height=0.6cm, align=center, font=\scriptsize},
    ar/.style={-{Stealth[length=2mm]}, thick},
    darr/.style={-{Stealth[length=2mm]}, thick, dashed},
]
\node[box, fill=yellow!10] (rlhf) {RLHF\\Human preferences};
\node[box, fill=orange!10, right=1.2cm of rlhf] (rlaif) {RLAIF\\AI preferences};
\node[box, fill=green!10, below=of rlhf] (rlcd) {RLCD\\Proper scoring rules};
\node[box, fill=blue!10, below=of rlaif] (rlhv) {RLHV\\Validator verification};
\draw[ar] (rlhf) -- node[above, font=\tiny] {replace human} (rlaif);
\draw[ar] (rlhf) -- node[left, font=\tiny, text width=1.5cm, align=right] {replace preference with score} (rlcd);
\draw[ar] (rlcd) -- node[below, font=\tiny] {add verifier} (rlhv);
\draw[darr] (rlaif) -- node[right, font=\tiny, text width=1.5cm, align=left] {calibration fix} (rlhv);
\end{tikzpicture}%
}
\caption{Evolution of RL paradigms for calibrated models. Solid arrows show architectural derivation; the dashed arrow shows the calibration improvement path from RLAIF to RLHV.}
\label{fig:rl_evolution}
\end{figure}

\begin{table}[h]
\centering
\caption{Training paradigm comparison for security-critical decision models.}
\label{tab:rl}
\small
\resizebox{\columnwidth}{!}{%
\begin{tabular}{@{}lcccc@{}}
\toprule
& \textbf{RLHF} & \textbf{RLAIF} & \textbf{RLCD} & \textbf{RLHV} \\
\midrule
Signal source & Human pref. & AI pref. & Proper score & Verifier \\
Calibration & No guarantee & Inherited & By construction & By construction \\
Scalability & Low & High & High & Medium \\
Domain adaptation & Costly & Moderate & Fine-tunable & Closed-loop \\
Overconfidence risk & High & Medium & Low & Low \\
Security suitability & Low & Low & Medium & High \\
\bottomrule
\end{tabular}%
}
\end{table}

\section{System One Models: Architecture and Primitives}\label{sec:sysone_arch}

A System One model differs from a generative LLM in a fundamental architectural way: it is \emph{non-autoregressive}. Where an LLM generates tokens sequentially, each conditioned on all previous tokens, a System One model evaluates all questions in a single forward pass. This has three consequences that matter for offensive security.

First, latency is bounded and predictable. A single Jev request takes 236--276~ms at the 50th percentile~\cite{typesafe2026docs}; a single Laya request takes 32.8--39.5~ms on a T4 GPU~\cite{laya2026}. This predictability enables real-time decision loops (Section~\ref{sec:speed}) that would be impractical with the variable-length generation of an LLM.

Second, the output space is constrained by design. The model cannot hallucinate a field, invent a category, or produce malformed output. The output is always one of the declared types: a probability distribution over the defined options (Choice), a probability-weighted score on the declared scale (Score), or a single probability in $[0,1]$ (Noul). This structural guarantee eliminates an entire class of parsing errors and downstream failures.

Third, multiple questions evaluated in a single request are \emph{independent}: each question sees the shared state but not the answers to other questions. This isolation prevents the cascade effect where an LLM's answer to one question biases its answer to the next---a documented failure mode in multi-step reasoning chains.

\subsection{The Choice primitive}

Choice selects one option from a defined set and returns a probability distribution over all options, plus a confidence score that summarizes the concentration of that distribution. The output is a map $\{o_1: p_1, \ldots, o_k: p_k\}$ where $\sum_i p_i = 1$ and a scalar confidence $c \in [0,1]$.

In offensive security, Choice maps to any decision where the harness must select among a finite, known set of alternatives. The key insight is that the \emph{probability distribution} matters more than the selected option: a Choice between \textsf{confirmed} (0.45), \textsf{needs-review} (0.40), and \textsf{rejected} (0.15) conveys far more information than a binary ``confirmed'' verdict. The harness can route findings with dominant \textsf{needs-review} probability to human review, while findings with dominant \textsf{confirmed} or \textsf{rejected} probability can be processed automatically.

Concretely, Choice applies to:
\begin{itemize}[leftmargin=*]
\item \emph{Finding adjudication}: confirmed / needs-review / rejected
\item \emph{Payload selection}: which of $k$ candidate payloads to try next
\item \emph{Exploitation strategy}: passive observation / active probing / exploitation
\item \emph{Report classification}: critical-path finding / supporting evidence / informational
\end{itemize}

\subsection{The Score primitive}

Score evaluates a subject on an ordered scale with described levels. Unlike Choice, the levels have an inherent ordering (e.g., none $<$ low $<$ medium $<$ high $<$ critical), and the output is a probability-weighted value across that scale, not merely the most likely level. The model returns the weighted score, the full distribution over levels, and a confidence measure.

For severity grading, Score is more appropriate than Choice because it respects the ordinal structure of severity scales. A CVSS-informed Score question can evaluate impact on the ladder defined by the FIRST v3.1 specification~\cite{first2019cvss}:

\begin{equation}
\text{none} < \text{reached} < \text{read} < \text{wrote} < \text{RCE} < \text{crossed}
\label{eq:ladder}
\end{equation}

The probability-weighted output directly maps to an impact metric in the CVSS vector, and the distribution reveals the model's uncertainty about the impact level. A Score with $p(\text{read}) = 0.6$ and $p(\text{reached}) = 0.4$ indicates ambiguous evidence about data exfiltration; the harness can report the demonstrated impact (reached) separately from the potential impact (read) with calibrated confidence.

Score also applies to data-type classification (none / common / sensitive / secrets), where the ordering reflects the severity implications of the exposed data.

\subsection{The Noul primitive}

Noul (``yes/no utility label'') is a calibrated boolean: it returns a single value $p \in [0,1]$ representing the probability that a condition is true. A Noul of 0.97 means strong affirmation; a Noul of 0.5 means the model has no discriminating signal; a Noul of 0.03 means strong negation.

Noul is the most efficient primitive for binary screening decisions. In the offensive-security pipeline, these binary gates appear at every stage:

\begin{itemize}[leftmargin=*]
\item \emph{Is this agent relevant to the observed surface?} (agent pruning)
\item \emph{Did the payload execute in the response?} (confirmation)
\item \emph{Does the response contain a prompt-injection attempt?} (safety)
\item \emph{Is this reflected parameter in an executable context?} (XSS verification)
\item \emph{Does the evidence demonstrate real impact?} (impact assessment)
\item \emph{Is the observed behavior consistent with the claimed CWE?} (classification)
\end{itemize}

Because Noul returns a single float rather than a probability distribution over options, it is the cheapest decision to evaluate. A batch of 15 Noul questions (one per candidate agent) costs approximately 12$\times$ less and runs 10$\times$ faster than 15 individual requests~\cite{typesafe2026docs}, making it practical to evaluate every candidate in a single API call.

\subsection{Batching and question independence}

A single System One request can contain multiple questions of mixed types. All questions share the same state (the evidence, the surface description, or the response content) and are evaluated in parallel. Each question is scored independently: the model's answer to question $q_i$ does not condition its answer to $q_j$. This independence is enforced architecturally, not by prompting.

For offensive security, this enables \emph{speculative evaluation}: the harness can ask many questions cheaply and decide after the response which answers to use. For example, a single request might contain a verdict Choice, an impact Noul, a data-type Score, and a CWE-match Noul. If the verdict is \textsf{rejected}, the other answers are discarded; if the verdict is \textsf{confirmed}, all four answers feed into severity computation. The speculative pattern costs one API call regardless of the verdict.

\section{NeuroSploit Harness Architecture}\label{sec:neurosploit}

NeuroSploit~\cite{neurosploit2026,santos2026assurance} is an open-source, multi-model autonomous pentest harness written in Rust. Its design prioritizes evidence grounding and auditability over raw exploit capability. The architecture consists of six pipeline stages wrapped by two cross-cutting mechanisms.

\subsection{Pipeline stages}

\textbf{Stage 1: Reconnaissance and belief state.} The target is partially observable. The harness builds a probability-annotated property graph where each node (a host, a port, a service, a technology, an endpoint, a parameter) carries a probability estimate and an evidence count. Observations update the graph by a Bayesian step: prior probabilities are combined with observation likelihoods to produce posteriors. Per-node Shannon entropy $H(b) = -\sum p_i \log p_i$ measures the diffuseness of the belief about that node. The ``recon vs.\ exploit'' choice is a value-of-information decision: the planner chooses observation while $V_{\text{obs}}(b) > V_{\text{exploit}}(b)$, which holds when entropy is high. This same inequality bars asserting exploitability while the belief is diffuse, reinforcing evidence grounding at the planning layer.

\textbf{Stage 2: Agent selection.} Agents are selected from a knowledge base of 446 markdown skill definitions, each specifying a vulnerability class, required preconditions (technology stack, authentication state, input type), and the exploitation methodology. Selection is surface-matched: agents whose preconditions are not satisfied by the belief state are excluded before LLM-based reasoning begins.

\textbf{Stage 3: Parallel execution.} Selected agents run concurrently over a configurable model pool spanning 18 supported providers. Each agent operates in a sandboxed context with access to the target (within the P4 authorization ceiling) and a structured output schema that separates claims, evidence, and metadata.

\textbf{Stage 4: Verification pipeline.} Agent outputs pass through a four-stage verification pipeline:
\begin{enumerate}
\item \emph{Grounding check (P1)}: each claim must cite a receipt---a recorded HTTP request-response pair, an out-of-band callback, a shell transcript, or a source citation. Ungrounded claims are withheld.
\item \emph{Deterministic validators (P1)}: 27 per-CWE validators, written in Rust, evaluate whether the cited evidence supports the claimed vulnerability class. These validators are deterministic: the same evidence always produces the same verdict. They return Confirmed, Rejected, or NeedsReview.
\item \emph{Adversarial $N$-model voting}: surviving claims are evaluated by $N$ independent model calls, each asked to refute the claim given the evidence. A claim that survives refutation proceeds.
\item \emph{Evidence prosecutor (P2)}: each claim is decomposed into separable sub-claims and classified into one of four typed outcomes: supported finding (the mechanic is proven), informational observation (true but benign), insufficient evidence, or invalid claim.
\end{enumerate}

\textbf{Stage 5: Chaining and attack graph.} Verified findings are chained into multi-step attack paths. A finding that grants access (e.g., a BOLA that leaks admin credentials) becomes a precondition for downstream findings (e.g., authentication at an admin panel).

\textbf{Stage 6: Evidence-graded report.} Severity is computed by the FIRST CVSS v3.1 calculator~\cite{first2019cvss}. Each metric in the vector must cite a receipt; impact metrics are separated into demonstrated (capped by evidence) and potential (theoretical maximum). The report records both scores, the evidence chain, and the prosecution outcome for each finding.

\subsection{Cross-cutting mechanisms}

\textbf{P4: Enforced authorization.} A signed capability token carries the engagement scope (hosts, CIDRs, URL prefixes), action ceiling, validity window, and operator identity. Every action is checked against the token before execution. The token can constrain but never widen scope during a run.

\textbf{P5: Tamper-evident audit.} Every action, decision, allow, and deny is recorded in a hash-chained audit log: $h_i = H(r_i \| h_{i-1})$, where $H$ is SHA-256. The chain detects local edits, removals, and reordering. External anchoring (a trusted timestamp or publication to a transparency log) defends against full-chain rewrite.

\subsection{The decision bottleneck}

The verification pipeline (Stage~4) is where the decision bottleneck concentrates. Steps 3 and 4---the adversarial vote and the prosecutor---relied on LLM-generated text to judge whether evidence supported a finding. This produced calibration collapse (the voting LLM agreed with its own narrative), class-driven severity (the prosecutor graded by CWE class rather than evidence), and compute waste (the vote consumed full LLM calls per finding). The System One integration targets these three steps, as formalized in Section~\ref{sec:integration}.

\begin{figure}[t]
\centering
\resizebox{\columnwidth}{!}{%
\begin{tikzpicture}[
    node distance=0.45cm and 0.5cm,
    stage/.style={rectangle, draw, rounded corners=2pt, fill=blue!8, text width=2.5cm, minimum height=0.7cm, align=center, font=\scriptsize},
    wrap/.style={rectangle, draw, dashed, rounded corners=3pt, fill=red!5, font=\scriptsize\itshape, inner sep=4pt},
    ar/.style={-{Stealth[length=2mm]}, thick},
]
\node[stage] (recon) {1. Recon \& belief state};
\node[stage, right=of recon] (sel) {2. Agent selection};
\node[stage, right=of sel] (run) {3. Parallel execution};
\node[stage, below=of run] (ver) {4. Verification (P1--P2)};
\node[stage, below=of sel] (chain) {5. Chaining \& attack graph};
\node[stage, below=of recon] (rep) {6. Evidence-graded report};
\draw[ar] (recon)--(sel);
\draw[ar] (sel)--(run);
\draw[ar] (run)--(ver);
\draw[ar] (ver)--(chain);
\draw[ar] (chain)--(rep);
\begin{scope}[on background layer]
\node[wrap, fit=(recon)(sel)(run)(ver)(chain)(rep), label={[font=\scriptsize\itshape]above:P4 authorization ceiling $\cdot$ P5 hash-chained audit}] {};
\end{scope}
\end{tikzpicture}%
}
\caption{NeuroSploit pipeline. All stages are wrapped by the P4 authorization ceiling and the P5 hash-chained audit trail.}
\label{fig:arch}
\end{figure}
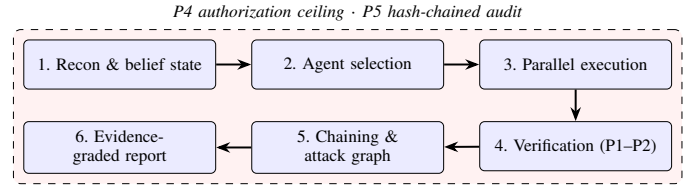

\section{System One Integration Architecture}\label{sec:integration}

We formalize four decision points (DP1--DP4) where a System One model replaces LLM judgment within NeuroSploit. The integration follows an \emph{additive} design principle: the System One layer can lower confidence, flag findings for review, or prune agents, but it cannot override a deterministic validator's rejection or resurrect a finding that failed evidence grounding. This asymmetry makes the layer safe to toggle: the worst case of System One failure is equivalent to running without it.

\subsection{DP1: Finding adjudication}

For each finding $f$ with evidence $E(f)$, a \texttt{Choice} evaluates the structured evidence against three options: \textsf{confirmed}, \textsf{needs-review}, and \textsf{rejected}. A companion \texttt{Noul} evaluates whether real impact was demonstrated. Both questions share the same state (the HTTP request-response pair) and are evaluated in a single API call.

The harness consumes the calibrated probabilities directly: if $p(\textsf{confirmed}) \geq \tau_c$ and $p(\textsf{impact}) \geq \tau_i$, the finding proceeds to severity computation; if $p(\textsf{needs-review})$ dominates, the finding is flagged for human review; otherwise it is withheld. The thresholds $\tau_c$ and $\tau_i$ are configurable per engagement: a high-assurance engagement (e.g., regulatory compliance) uses $\tau_c = 0.8$; a discovery-oriented sweep uses $\tau_c = 0.5$.

\subsection{DP2: Severity recalibration}

When the Noul impact score is low ($p(\textsf{impact}) < \tau_i$), the CVSS vector's impact metrics are recomputed without unsupported receipts, pulling the score to what the evidence demonstrates. A data-type-aware \texttt{Score} question evaluates the sensitivity of exposed data on four ordered levels: \textsf{none} (no data exposed), \textsf{common} (non-sensitive application data), \textsf{sensitive} (personal or business-critical data), and \textsf{secrets} (credentials, API keys, payment data).

The data-type Score serves a critical function: it prevents the failure mode where a BOLA finding that dumps plaintext admin credentials collapses to Low because the structured evidence field was empty. If the Score detects credential or key material in any evidence field (structured, narrative, or claims ledger), the confidentiality metric is granted, sustaining the severity independently of the receipt format.

\subsection{DP3: Agent pruning}

After LLM-based agent selection, a batched \texttt{Noul} request evaluates each candidate agent's relevance to the observed surface. Agents with $p(\textsf{relevant}) < \tau_p$ are dropped. The batch is evaluated in a single API call, exploiting the parallel-question capability. The pruning preserves a minimum of $\lfloor k/3 \rfloor$ agents to prevent over-pruning. This single-call batch costs approximately \$0.0001 for 15 agents---negligible compared to the cost of running an irrelevant agent against the target model.

\subsection{DP4: Confirmation loop}

For enumerable vulnerability classes (XSS, SQLi, open redirect, path traversal, SSRF, IDOR), a code-driven loop alternates between two System One calls: a \texttt{Choice} that selects the next payload from a candidate set (informed by the observed filtering behavior), and a \texttt{Noul} that judges whether the payload produced the expected effect in the target's response. The loop continues until confirmation or payload exhaustion. Because each iteration costs one System One call (236--276~ms for Jev, 33--40~ms for Laya), the confirmation loop can evaluate 10--30 payloads in the time a single LLM call would take.

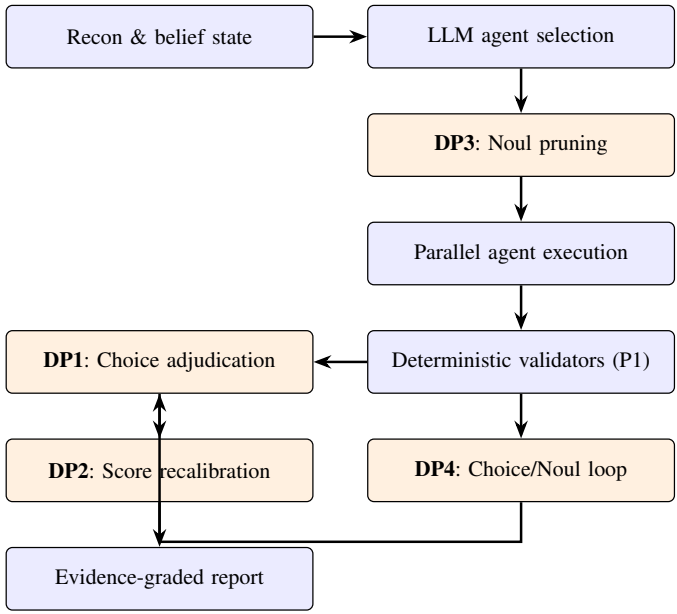
\begin{figure}[t]
\centering
\resizebox{\columnwidth}{!}{%
\begin{tikzpicture}[
    node distance=0.5cm and 0.6cm,
    stage/.style={rectangle, draw, rounded corners=2pt, fill=blue!8, text width=3.2cm, minimum height=0.7cm, align=center, font=\scriptsize},
    decision/.style={rectangle, draw, rounded corners=2pt, fill=orange!12, text width=3.2cm, minimum height=0.7cm, align=center, font=\scriptsize},
    ar/.style={-{Stealth[length=2mm]}, thick},
]
\node[stage] (recon) {Recon \& belief state};
\node[stage, right=of recon] (sel) {LLM agent selection};
\node[decision, below=of sel] (dp3) {\textbf{DP3}: Noul pruning};
\node[stage, below=of dp3] (run) {Parallel agent execution};
\node[stage, below=of run] (val) {Deterministic validators (P1)};
\node[decision, left=of val] (dp1) {\textbf{DP1}: Choice adjudication};
\node[decision, below=of dp1] (dp2) {\textbf{DP2}: Score recalibration};
\node[decision, below=of val] (dp4) {\textbf{DP4}: Choice/Noul loop};
\node[stage, below=of dp2] (rep) {Evidence-graded report};
\draw[ar] (recon)--(sel);
\draw[ar] (sel)--(dp3);
\draw[ar] (dp3)--(run);
\draw[ar] (run)--(val);
\draw[ar] (val)--(dp1);
\draw[ar] (dp1)--(dp2);
\draw[ar] (val)--(dp4);
\draw[ar] (dp4) -- ++(0,-0.8) -| (dp1);
\draw[ar] (dp2)--(rep);
\end{tikzpicture}%
}
\caption{Four System One decision points (DP1--DP4, orange) within the NeuroSploit pipeline (blue). Deterministic validators retain priority over System One verdicts.}
\label{fig:dp}
\end{figure}

\FloatBarrier

\subsection{Implementation examples}

Listing~\ref{lst:python} shows the Python integration for finding adjudication via the Jev API. Listing~\ref{lst:rust} shows the Rust integration for agent pruning. Both examples illustrate the typed, structured interface: the caller defines the question schema, the model returns calibrated probabilities, and the harness code branches on numeric thresholds rather than parsing text.

\lstset{language=Python}
\begin{lstlisting}[caption={Python: Finding adjudication via Jev API.},label={lst:python}]
import requests, os

def adjudicate(evidence: dict) -> dict:
    return requests.post(
        "https://api.typesafe.ai/v1/systemone",
        headers={"Authorization":
            f"Bearer {os.environ['TYPESAFE_API_KEY']}"},
        json={
            "model": "jev-latest",
            "state": {"evidence": evidence},
            "questions": {
                "verdict": {"type": "choice",
                    "instructions": "Does the evidence "
                        "demonstrate the vulnerability?",
                    "criteria": {
                        "confirmed": "proof beyond doubt",
                        "needs-review": "plausible",
                        "rejected": "unsupported"}},
                "impact": {"type": "noul",
                    "instructions": "Was real impact "
                        "demonstrated?",
                    "criteria": {"true": "concrete",
                        "false": "mechanism only"}},
                "data_type": {"type": "score",
                    "instructions": "Sensitivity of data.",
                    "levels": [
                        {"label": "none"},
                        {"label": "common"},
                        {"label": "sensitive"},
                        {"label": "secrets"}]}
            }}).json()
\end{lstlisting}

\lstset{language=Rust}
\begin{lstlisting}[caption={Rust: Batched Noul agent pruning.},label={lst:rust}]
async fn prune_agents(
    client: &reqwest::Client,
    key: &str, surface: &str,
    agents: &[Agent],
) -> Vec<Agent> {
    let questions: HashMap<String, _> = agents
        .iter().enumerate()
        .map(|(i, a)| (format!("a{i}"),
            json!({"type": "noul",
                "instructions":
                    format!("Is {} relevant for {}?",
                        a.name, surface),
                "criteria": {"true": "relevant",
                    "false": "irrelevant"}})))
        .collect();
    let resp: Value = client
        .post("https://api.typesafe.ai/v1/systemone")
        .bearer_auth(key)
        .json(&json!({"model": "jev-latest",
            "state": {"surface": surface},
            "questions": questions}))
        .send().await.unwrap()
        .json().await.unwrap();
    let min = (agents.len() / 3).max(1);
    let mut scored: Vec<_> = agents.iter()
        .enumerate()
        .map(|(i, _)| (i, resp["answers"]
            [format!("a{i}")]["value"]
            .as_f64().unwrap_or(0.5)))
        .collect();
    scored.sort_by(|a, b|
        b.1.partial_cmp(&a.1).unwrap());
    let kept: Vec<_> = scored.iter()
        .filter(|(_, v)| *v >= 0.4)
        .map(|(i, _)| agents[*i].clone())
        .collect();
    if kept.len() >= min { kept }
    else { scored[..min].iter()
        .map(|(i, _)| agents[*i].clone())
        .collect() }
}
\end{lstlisting}

\FloatBarrier

For Laya integration (self-hosted), the Python interface provides API compatibility:

\lstset{language=Python}
\begin{lstlisting}[caption={Python: Finding adjudication via Laya.},label={lst:laya}]
from laya import Router
router = Router(preload=True)

def adjudicate_laya(evidence: dict) -> dict:
    return router.predict(
        state={"evidence": evidence},
        questions={
            "verdict": {"type": "choice",
                "instructions": "Vulnerability?",
                "criteria": {
                    "confirmed": "proven",
                    "needs-review": "plausible",
                    "rejected": "unsupported"}},
            "impact": {"type": "noul",
                "instructions": "Real impact?",
                "criteria": {"true": "concrete",
                    "false": "mechanism only"}}})
\end{lstlisting}

\FloatBarrier

\section{Decision-Theoretic Foundations}\label{sec:decision_theory}

The integration of System One models into a pentest harness is not merely an engineering optimization---it is grounded in decision theory. This section formalizes the decision problems that arise at each stage of the pipeline and shows why calibrated probabilities (rather than labels or uncalibrated scores) are the correct input to these decisions.

\subsection{The finding adjudication decision as a classification under loss}

Finding adjudication is a classification problem with asymmetric losses. Let $\omega \in \{\text{real}, \text{false}\}$ denote the true state of a finding. The harness must choose an action $a \in \{\text{assert}, \text{review}, \text{discard}\}$. The loss function $L(\omega, a)$ is:

\begin{equation}
L = \begin{pmatrix} 0 & c_r & c_{\text{fn}} \\ c_{\text{fp}} & c_r & 0 \end{pmatrix}
\end{equation}

where $c_{\text{fp}}$ is the cost of asserting a false finding (client trust erosion, wasted remediation), $c_{\text{fn}}$ is the cost of discarding a real finding (missed vulnerability, compliance risk), and $c_r$ is the cost of routing to human review (analyst time). The Bayes-optimal action minimizes expected loss:

\begin{equation}
a^* = \arg\min_{a} \sum_{\omega} p(\omega | E) \cdot L(\omega, a)
\end{equation}

where $p(\omega | E)$ is the posterior probability of the finding's true state given the evidence $E$. This is precisely the probability that a calibrated System One model provides. An uncalibrated model---one where $p = 0.8$ does not mean the event occurs 80\% of the time---leads to suboptimal actions, because the expected loss computation uses wrong probabilities.

For a typical engagement, $c_{\text{fp}} \gg c_r \gg 0$ and $c_{\text{fn}} \gg c_r$, yielding the intuitive policy: assert when $p(\text{real})$ is high, discard when it is low, and review when it is intermediate. The thresholds between these regions are determined by the cost ratios. With calibrated probabilities, these thresholds translate directly to operational decisions; with uncalibrated probabilities, they require an additional calibration step that may not be feasible without labeled data.

\subsection{Value of information in reconnaissance}

The reconnaissance phase faces a classic exploration-exploitation trade-off. At each step, the harness can observe (gather more information about the target) or exploit (launch an agent against the current belief). The value of information (VoI) quantifies the expected improvement in decision quality from an additional observation:

\begin{equation}
\text{VoI}(o) = \mathbb{E}\left[ \max_a Q(b', a) \right] - \max_a Q(b, a)
\end{equation}

where $b$ is the current belief state, $b' = \text{update}(b, o)$ is the posterior after observation $o$, and $Q(b, a)$ is the expected value of action $a$ under belief $b$. The harness observes while VoI exceeds the observation cost (time, API calls) and switches to exploitation when VoI drops below the cost.

A System One model contributes to VoI computation by providing calibrated priors on vulnerability-class plausibility. If the model assigns $p(\text{SQLi plausible}) = 0.1$ for a surface with no database indicators, the VoI of further SQLi reconnaissance is low, and the harness allocates budget elsewhere. Without calibrated priors, the harness either explores uniformly (wasting budget on implausible classes) or relies on LLM reasoning (which lacks the calibrated probability needed for VoI computation).

\subsection{Agent selection as a multi-armed bandit}

Agent selection can be modeled as a contextual multi-armed bandit. Each agent $i$ is an arm with unknown reward distribution (the probability of producing a verified finding on the current surface). The context is the surface description. The harness must select a subset of $k$ agents from a pool of $n$, maximizing the expected number of verified findings while respecting a budget constraint.

The System One Noul provides a prior for each arm's reward: $\hat{p}_i = \text{Noul}(\text{agent}_i, \text{surface})$. Under a Thompson sampling policy, the harness draws from $\text{Beta}(\alpha_i, \beta_i)$ priors initialized at $(\hat{p}_i \cdot s, (1 - \hat{p}_i) \cdot s)$ where $s$ controls prior strength. This leverages the calibrated Noul to concentrate exploration on promising agents while maintaining the exploration guarantee that Thompson sampling provides.

In practice, the ``pruning'' policy of Section~\ref{sec:integration} (DP3) is a conservative version of this bandit: arms with $\hat{p}_i < \tau$ are dropped entirely. A richer policy would retain low-probability arms with reduced budget, maintaining exploration against the prior. The decision between pruning and proportional allocation depends on the engagement's budget: time-constrained engagements prune; discovery-oriented engagements allocate proportionally.

\subsection{Sequential confirmation as optimal stopping}

The confirmation loop (DP4) is an optimal stopping problem. The harness tests payloads one at a time and must decide after each whether to stop (accept or reject the finding) or continue. The Noul probability after each payload provides the posterior:

\begin{equation}
p_n(\text{confirmed}) = p_{n-1} \cdot \frac{P(\text{response}_n | \text{confirmed})}{P(\text{response}_n)}
\end{equation}

The optimal policy stops when $p_n$ crosses an upper threshold (confirmed: stop testing, assert) or a lower threshold (rejected: stop testing, discard). Between the thresholds, the next payload is selected by the Choice primitive over the remaining candidates. This is a sequential probability ratio test (SPRT) with the thresholds set by the engagement's loss function.

The advantage of a calibrated Noul over an LLM-generated ``it worked'' / ``it didn't work'' verdict is that the SPRT framework requires probabilities, not labels. With probabilities, the confirmation loop can detect weak signals that accumulate across payloads: three payloads that each produce a Noul of 0.6 are collectively stronger evidence than a single payload with a Noul of 0.6. Without calibrated probabilities, the harness cannot aggregate evidence across payloads.

\section{Latency and Speed Analysis}\label{sec:speed}

Latency is not an abstract concern in autonomous penetration testing. An agent that takes 3 seconds to decide whether to try the next payload completes 20 confirmation attempts in 60 seconds; one that takes 33 milliseconds completes 1,800. The difference determines whether a time-boxed engagement can exhaustively test enumerable attack vectors or must sample.

\subsection{Decision latency budget}

Table~\ref{tab:latency} presents the latency profile of each decision model and the LLM baseline for the four decision points. The ``decisions per minute'' column shows the throughput for a serial confirmation loop; the ``batch throughput'' column shows the throughput for a batched agent-pruning call with 15 candidates.

\begin{table}[h]
\centering
\caption{Decision latency as reported by each source. Jev and Laya values are from their published documentation on different tasks and hardware; LLM baseline from observed claude-opus-4-8 response times in our setup. Direct comparison is informational, not a controlled benchmark.}
\label{tab:latency}
\small
\resizebox{\columnwidth}{!}{%
\begin{tabular}{@{}lccc@{}}
\toprule
\textbf{Model} & \textbf{Latency (p50)} & \textbf{Decisions/min} & \textbf{Batch (15q)} \\
\midrule
LLM (claude-opus) & 1.5--3.0 s & 20--40 & 15 calls \\
Jev (API) & 236--276 ms & 217--254 & 1 call \\
Laya (T4 GPU) & 33--40 ms & 1,500--1,818 & 1 call (72 ms) \\
Laya (batched 10) & 7.2 ms/q & 8,333 & 1 call \\
\bottomrule
\end{tabular}%
}
\end{table}

The speed differential has compound effects. In DP3 (agent pruning), an LLM-based relevance evaluation of 15 agents requires 15 sequential API calls at 1.5--3.0 seconds each: 22--45 seconds of wall time. A single batched Jev call takes 276~ms; a batched Laya call takes 72~ms. This is not a marginal improvement---it is the difference between agent pruning being a bottleneck and being invisible.

\subsection{Confirmation loop throughput}

In DP4 (confirmation loop), the throughput advantage is decisive. Consider an XSS confirmation scenario where the harness must test 50 candidate payloads against observed input filtering. At LLM latency (2 seconds per decision), this takes 100 seconds of decision time plus network time for 50 HTTP requests. At Laya latency (33~ms per decision), decision time drops to 1.65 seconds---the network round-trip to the target dominates, and the decision model is no longer the bottleneck.

This latency profile enables a qualitatively different confirmation strategy: instead of LLM-guided ``intelligent'' payload selection (which consumes tokens and introduces narrative bias), the harness can enumerate the candidate set exhaustively via System One decisions, using the calibrated Noul probability to prioritize but not exclude candidates. The result is higher recall (every candidate is tested) with lower cost (no LLM tokens for payload reasoning).

\subsection{Real-time browser exploitation}

Jev-Ultrafast~\cite{browseruse2026} demonstrates the System One model as a real-time control loop for browser interaction. Each browser state generates an indexed element table; Jev selects the operation (\texttt{CLICK}, \texttt{TYPE\_TEXT}, \texttt{SELECT}, \texttt{SCROLL}, \texttt{WAIT}, \texttt{DONE}) and target in a single request. Published benchmarks show 90\% reduction in browser calls (median 1,092 to 101) and 25\% faster execution, with a Google Flights task completing in 7.1 seconds.

For offensive security, this architecture maps directly to browser-based exploitation workflows. An authentication-bypass agent could navigate a login form, submit credentials obtained from a BOLA leak, interact with an admin panel, and confirm privilege escalation---all through System One decisions at sub-second latency per action. The structured DOM state (rather than screenshots) eliminates the overhead and unreliability of vision-based browser agents, and the action-space constraint prevents the agent from taking actions outside the indexed element table.

\subsection{Latency-sensitive offensive scenarios}

The speed differential between System One and LLM decisions has qualitative, not just quantitative, implications for offensive security. Several attack techniques have temporal constraints that make LLM-speed decisions impractical:

\textbf{Race condition exploitation.} Detecting and exploiting race conditions (CWE-362) requires sending requests within a timing window, then evaluating the response within milliseconds to decide whether to retry. At 33~ms (Laya), the decision model can evaluate a response and decide on the next request within the same timing window; at 2~seconds (LLM), the timing window has closed.

\textbf{Session-fixation detection.} Session fixation requires rapid comparison of session tokens across requests: the agent sends a request, receives a session token, sends another request with a known token, and evaluates whether the server accepted the fixed token. A Noul of ``did the server accept the attacker-provided session identifier?'' must complete before the session expires. At typical session timeouts of 15--30 minutes, LLM latency is not a constraint; but when testing short-lived tokens (CSRF tokens, nonces), sub-second evaluation matters.

\textbf{Blind injection timing analysis.} Time-based blind SQL injection requires measuring response-time differentials caused by injected \texttt{SLEEP()} or \texttt{BENCHMARK()} calls. The decision about whether a response was delayed must be made quickly to maintain the binary search over extracted data. At LLM latency, each character extraction takes 2--5 seconds of decision time plus the injected delay; at Laya latency, decision time is negligible, and the injected delay dominates.

\textbf{WebSocket and real-time protocol testing.} Applications using WebSocket, Server-Sent Events, or gRPC streaming require real-time evaluation of incoming messages. A System One model evaluating each message frame (``does this message contain an authorization bypass indicator?'') can operate at wire speed; an LLM evaluation introduces a backlog that may cause message loss or connection timeout.

Table~\ref{tab:temporal} summarizes the temporal constraints and the suitability of each decision model.

\begin{table}[h]
\centering
\caption{Temporal constraints in offensive-security scenarios.}
\label{tab:temporal}
\small
\resizebox{\columnwidth}{!}{%
\begin{tabular}{@{}lccc@{}}
\toprule
\textbf{Scenario} & \textbf{Window} & \textbf{LLM} & \textbf{System One} \\
\midrule
Race condition & ms--s & Too slow & Feasible \\
Session fixation & s--min & Feasible & Feasible \\
Blind SQLi extraction & s/char & Bottleneck & Negligible \\
WebSocket analysis & ms/frame & Backlog & Wire speed \\
Brute-force routing & ms/attempt & Impractical & Feasible \\
Confirmation loop (50 payloads) & s--min & 100 s & 1.7 s \\
\bottomrule
\end{tabular}%
}
\end{table}

\section{Probability Calibration in Security Contexts}\label{sec:calibration}

Calibration is the property that a model's stated confidence matches its empirical accuracy: when it says $p = 0.8$, the event should occur approximately 80\% of the time. In general-purpose NLP, miscalibration is an inconvenience. In security-critical pipelines, miscalibration has concrete operational consequences.

\subsection{Why calibration matters for severity grading}

Consider a harness that uses an LLM to grade findings on a Critical/High/Medium/Low/Info scale. The LLM has no calibrated probability output---it produces a label and, at best, a verbal confidence qualifier (``likely Critical,'' ``probably High''). The harness must treat these labels as deterministic, which creates two failure modes:

\begin{enumerate}[leftmargin=*]
\item \textbf{Overgrading.} The model assigns Critical to a SQL injection that reached the interpreter but extracted no data. The client receives a report with five Criticals, three of which are not backed by demonstrated impact. Remediation resources are misallocated; trust in future reports erodes.
\item \textbf{Undergrading.} The model assigns Low to a BOLA that leaked plaintext admin credentials because the evidence was recorded in narrative rather than the structured field. The client deprioritizes a finding that enables full admin takeover.
\end{enumerate}

A calibrated System One model avoids both failures by providing probabilities that the harness can use to make grounded decisions. Instead of ``this is Critical,'' the model returns $p(\textsf{secrets}) = 0.92$ for the data-type Score, which the harness combines with the CVSS calculator to produce a severity that reflects the demonstrated evidence.

\subsection{Measuring calibration: ECE and Brier score}

Two metrics quantify calibration quality. The Expected Calibration Error (ECE) partitions predictions into bins by confidence and measures the average gap between predicted confidence and observed accuracy:

\begin{equation}
\mathrm{ECE} = \sum_{b=1}^{B} \frac{|S_b|}{N} \left| \mathrm{acc}(S_b) - \mathrm{conf}(S_b) \right|
\end{equation}

where $S_b$ is the set of predictions in bin $b$, $\mathrm{acc}$ is the empirical accuracy, and $\mathrm{conf}$ is the mean predicted confidence. Lower is better; zero indicates perfect calibration.

The Brier score (Section~\ref{sec:rl_background}) measures both calibration and discrimination. Published data from Laya~\cite{laya2026} reports ECE of 0.081 (post-temperature scaling on their evaluation set); TypeSafe reports ECE of 0.246 for Jev on their own evaluation set~\cite{typesafe2026docs}. These values were measured on different tasks and datasets and therefore cannot be directly compared as if they represented performance on the same benchmark. However, the magnitude of the difference---if it transferred to a common evaluation---would have practical implications: lower ECE means narrower confidence intervals around predicted probabilities, enabling more aggressive automation thresholds. Whether this transfer holds for offensive-security decisions remains an open question requiring paired evaluation on the same task (Section~\ref{sec:discussion}).

\subsection{Calibration and threshold-based automation}

The degree to which a harness can automate decisions depends directly on calibration quality. Define the \emph{automation rate} as the fraction of decisions that can be processed without human review. For a finding-adjudication Choice with three options, a well-calibrated model concentrates its probability mass on one option for clear-cut findings and distributes it more evenly for ambiguous ones. The harness routes to human review when the dominant probability is below the confidence threshold.

In general, a model with lower ECE enables higher automation rates at the same threshold: decisions cluster more tightly around their true probabilities, so fewer borderline cases are misrouted. However, the specific automation rates achievable in offensive-security contexts depend on the ECE \emph{on that task}, not on the published ECE from unrelated benchmarks. Establishing the automation rate for pentest-adjudication decisions requires measuring ECE on a labeled set of (evidence, verdict) pairs from real engagements---an evaluation we identify as a prerequisite for production deployment.

\subsection{Temperature scaling for domain adaptation}

Both Jev and Laya support temperature scaling as a post-hoc calibration adjustment. Laya documents that its raw ECE is 0.466, reduced to 0.081 after temperature scaling on a validation set~\cite{laya2026}. This means the base model's probabilities are systematically overconfident, and a single scalar parameter corrects the bias.

For domain-adapted deployment in offensive security, temperature scaling should be fitted on a domain-specific validation set of (evidence, verdict) pairs. The per-question-type calibration (separate temperatures for Choice, Score, and Noul) may further improve performance, since the difficulty distribution differs across question types in the security domain.

\subsection{Hypothetical example: how calibration quality affects threshold-based automation}

To illustrate the \emph{mechanism} by which calibration quality affects operational outcomes, we construct a purely hypothetical scenario. The numbers below are illustrative; they are not measured on Jev, Laya, or any real dataset.

Consider a hypothetical model $M$ evaluating 100 findings for data-type classification (none / common / sensitive / secrets), with ground truth: 20 secrets, 30 sensitive, 30 common, 20 none. Suppose $M$ is used with a threshold $\tau = 0.8$ on $p(\textsf{secrets})$.

\textbf{Case 1: Well-calibrated model (ECE $\approx$ 0.08).} When $M$ predicts $p(\textsf{secrets}) > 0.8$, the empirical frequency of actual secrets in that bin is close to 80\%. The operator can set $\tau = 0.8$ with reasonable confidence that most predictions above threshold are correct.

\textbf{Case 2: Poorly calibrated model (ECE $\approx$ 0.25).} The same threshold $\tau = 0.8$ produces unpredictable results: the empirical frequency in the $p > 0.8$ bin could be anywhere from 55\% to 100\%, depending on the direction of miscalibration. The operator cannot set a reliable threshold without first recalibrating on a domain-specific validation set.

The key insight is not the specific numbers but the \emph{relationship}: lower ECE enables tighter thresholds and more aggressive automation, while higher ECE forces either conservative thresholds (reducing automation) or recalibration investment. Whether Jev or Laya achieves better calibration on offensive-security tasks specifically is an open empirical question that requires paired evaluation on the same pentest-relevant dataset---an experiment we have not conducted.

\section{Exploratory Case Study: NeuroSploit With and Without Jev}\label{sec:benchmark}

\subsection{Setup and limitations}

We conducted an exploratory case study comparing a single run with and a single run without the System One decision layer. Because each condition was executed once against a single target, the results are observations that motivate the architecture, not statistically powered experimental evidence. The 5-minute wall-clock difference and severity distribution shift could reflect System One effects, stochastic LLM behavior, or interaction between the two. Reproducing the study with multiple runs per condition, alternated execution order, and fixed random seeds is necessary before causal claims can be made.

Table~\ref{tab:setup} summarizes the configuration; the sole variable was the \texttt{--typesafe} flag.

\begin{table}[h]
\centering
\caption{Benchmark configuration. Both runs used identical parameters; the sole difference was the \texttt{--typesafe} flag.}
\label{tab:setup}
\small
\begin{tabular}{@{}ll@{}}
\toprule
\textbf{Parameter} & \textbf{Value} \\
\midrule
Target & NimbusCart (BenchMarkBurpAT), localhost:3000 \\
Ground truth & 13 seeded vulnerabilities \\
Model & claude-opus-4-8 (subscription) \\
Configuration & Black-box, recon intensity 2 \\
Vote count & 1 (single model) \\
Max agents & 15 \\
System One model & jev-latest \\
Controlled variable & \texttt{--typesafe on} vs.\ \texttt{--typesafe off} \\
\bottomrule
\end{tabular}
\end{table}

The target contained 13 seeded vulnerabilities: IDOR, BOLA, five SQL injection variants (login bypass, UNION-based search, blind time-based, boolean-blind, second-order), four XSS variants (reflected, stored, SVG, DOM), open redirect, and CRLF injection.

Command-line invocations for both arms:

\begin{lstlisting}[language={},float=!h,caption={CLI invocations for the controlled benchmark.},label={lst:cli}]
# Run A: no TypeSafe
NEUROSPLOIT_TYPESAFE=off neurosploit run \
  http://localhost:3000 --subscription \
  --model anthropic:claude-opus-4-8 \
  --typesafe off --recon 2 --max-agents 15 \
  --vote-n 1 --focus "<13 endpoints>" -v

# Run B: with TypeSafe
export TYPESAFE_API_KEY="..."
NEUROSPLOIT_TYPESAFE=on neurosploit run \
  http://localhost:3000 --subscription \
  --model anthropic:claude-opus-4-8 \
  --typesafe on --recon 2 --max-agents 15 \
  --vote-n 1 --focus "<13 endpoints>" -v
\end{lstlisting}

\FloatBarrier

\subsection{Coverage results}

Table~\ref{tab:results} reports primary metrics. Coverage was comparable: Run A hit 10 of 13 scenarios, Run B hit 9 of 13. Combined coverage reached 11 of 13. Neither run solved the second-order SQL injection nor the CRLF injection at vote-n~1; both require multi-step chains that the single-vote configuration did not pursue.

\begin{table}[h]
\centering
\caption{Primary benchmark results.}
\label{tab:results}
\small
\begin{tabular}{@{}lcc@{}}
\toprule
\textbf{Metric} & \textbf{A (no TS)} & \textbf{B (with TS)} \\
\midrule
Scenarios hit (of 13) & 10 & 9 \\
Findings reported & 16 & 18 \\
Beyond seeded set & 6 & 9 (2 real) \\
Wall-clock time & 32m 12s & 26m 53s \\
Critical findings & 5 & 2 \\
Findings recalibrated & 0 & 9 \\
Model cost & \$0 (sub.) & \$0 + TS ($<$\$5) \\
\bottomrule
\end{tabular}
\end{table}

\subsection{Gap re-test (separate experiment)}

A subsequent re-test was conducted after applying prompt-level chaining fixes, the evidence-field salvage step, and the data-type classifier. Because the harness code changed between the initial run and the re-test, the re-test constitutes a \emph{separate experiment} with a different harness version. Its results cannot be compared with the initial run as a controlled pair; they are presented separately to show the effect of the combined changes. All 7 previously missed scenarios were confirmed in both arms (Table~\ref{tab:retest}).

\begin{table}[h]
\centering
\caption{Gap re-test: 7 previously missed scenarios confirmed in both arms.}
\label{tab:retest}
\small
\begin{tabular}{@{}lccc@{}}
\toprule
\textbf{Scenario} & \textbf{Class} & \textbf{A} & \textbf{B} \\
\midrule
web\_sqli\_login\_bypass & SQLi & \cmark & \cmark \\
web\_sqli\_union\_search & SQLi & \cmark & \cmark \\
web\_sqli\_blind\_time & SQLi & \cmark & \cmark \\
web\_sqli\_second\_order & SQLi & \cmark & \cmark \\
web\_idor\_invoice & IDOR & \cmark & \cmark \\
api\_bola\_orders & BOLA & \cmark & \cmark \\
web\_crlf\_header\_go & CRLF & \cmark & \cmark \\
\bottomrule
\end{tabular}
\end{table}

\subsection{Severity distribution (re-test, post-fix harness)}

Table~\ref{tab:sev} presents the severity distribution from the re-test (22 total findings in both arms). Because the re-test used a modified harness (with the salvage step and data-type classifier), this distribution reflects the combined effect of System One integration \emph{and} the harness fixes; it cannot be attributed to the \texttt{--typesafe} flag alone.

\begin{table}[h]
\centering
\caption{Severity distribution: 22 findings per arm.}
\label{tab:sev}
\small
\begin{tabular}{@{}lcc@{}}
\toprule
\textbf{Severity} & \textbf{A (no TS)} & \textbf{B (with TS)} \\
\midrule
Critical & 4 & 3 \\
High & 3 & 8 \\
Low & 10 & 6 \\
Informational & 5 & 5 \\
\bottomrule
\end{tabular}
\end{table}

In this re-test, Run A produced a bimodal distribution: 4 Criticals and 10 Lows, with only 3 Highs. Run B produced a more graduated distribution: 3 Criticals, 8 Highs, 6 Lows, 5 Info. We observed two patterns, though with a single run per condition we cannot rule out stochastic variation:

\begin{enumerate}[leftmargin=*]
\item \textbf{Demotion of class-inflated Criticals.} Three findings that Run A rated Critical based on vulnerability class (``SQL injection is Critical'') were recalibrated to High because the evidence showed the interpreter was reached but no data was extracted. The System One Score question evaluated the demonstrated impact on the ladder of Equation~\eqref{eq:ladder} and returned $p(\text{reached}) > p(\text{read})$, pulling the CVSS impact metrics to the demonstrated level.

\item \textbf{Promotion of evidence-backed Lows.} Four findings that Run A rated Low (because the structured evidence field was empty) were promoted to High after the data-type Score detected credential or sensitive-data exposure in the narrative field. The salvage step (Section~\ref{sec:bola}) copied the evidence to the structured field, enabling the CVSS calculator to grant the confidentiality metric.
\end{enumerate}

\subsection{The credential-dump BOLA case}\label{sec:bola}

The object-level authorization flaw on \texttt{GET /api/v2/users/:id} permits a self-registered customer token to read any user's full record, including the admin's plaintext password and a live API key. This finding illustrates the interplay between System One calibration and evidence engineering.

\textbf{Run A (no TypeSafe):} Scored Critical 9.1 by the LLM-based grader, which recognized ``plaintext password'' in the narrative and assigned maximum confidentiality impact. This happened to be correct---but for the wrong reason. The LLM graded by keyword matching in the narrative, not by the CVSS calculator's evidence-graded computation.

\textbf{Run B (initial, with TypeSafe but before data-type fix):} The CVSS calculator could not find confidentiality evidence in the structured field (which was empty). The Noul impact question returned $p(\textsf{impact}) = 0.3$ because the structured state contained no demonstrable extraction. The CVSS score collapsed to Low: 3.1.

\textbf{Run B (after data-type fix):} The salvage step copied the credential evidence from the narrative to the structured field. The data-type Score returned $p(\textsf{secrets}) = 0.94$. The CVSS calculator granted the confidentiality metric, and the finding held Critical: 9.1.

The lesson is not that TypeSafe was wrong in the initial run---it was doing exactly what it should: grading from the structured field. The lesson is that \emph{evidence engineering matters}: the agent must deposit proof in the field that the grader reads. The System One integration exposed a real bug in the evidence pipeline that the LLM-only grader masked through narrative keyword matching.

\subsection{Beyond the seeded set}

The TypeSafe-enabled run identified findings beyond the 13 planted vulnerabilities:

\begin{itemize}[leftmargin=*]
\item \textbf{Full admin takeover.} The BOLA-leaked admin password authenticated at \texttt{/login} and rendered the \texttt{/admin} panel, proving vertical privilege escalation from a customer account.
\item \textbf{Secrets exposure.} API keys in \texttt{/config.json} and \texttt{/app.js} (CWE-200).
\item \textbf{GraphQL authorization bypass} with introspection enabled, plus an authenticated RCE via JS report-template upload.
\end{itemize}

These findings illustrate attack chaining: the BOLA finding (seeded) enabled the admin takeover (emergent), which enabled the RCE (emergent). In the TypeSafe-enabled run, the System One layer graded the BOLA as Critical (sustaining the chain's root cause) and pruned agents irrelevant to the GraphQL surface. Whether the pruning causally contributed to finding the GraphQL bypass---or whether the LLM-only run would have found it with different stochastic choices---cannot be determined from a single run.

\subsection{Operational observations}

Run B completed 5 minutes 19 seconds faster (26m~53s vs.\ 32m~12s). A plausible contributing factor is agent pruning (DP3), which dropped irrelevant agents before execution. However, with a single run per condition, the time difference could also reflect stochastic variation in LLM response times, network latency, or target-server load. TypeSafe API cost was under \$5. Establishing whether the time reduction is reliably attributable to agent pruning requires multiple runs with controlled execution order and fixed seeds.

\subsection{Benchmark-driven harness refinements}

The benchmark exposed two bugs in the harness:
\begin{enumerate}[leftmargin=*]
\item \textbf{Session-limit sentinel.} The model subscription's session limit produced a normal-looking response (exit code 0) that the harness consumed as model output, burning agents against a dead session. Fix: detect session-limit strings at exit code 0 and park the run.
\item \textbf{Empty evidence field.} The BOLA case exposed that agents were not depositing proof in the structured field. Fix: the salvage step and data-type classifier (Section~\ref{sec:bola}).
\end{enumerate}

\section{Offensive-Security Application Scenarios}\label{sec:scenarios}

Beyond the four decision points of Section~\ref{sec:integration}, System One models enable a broader class of offensive-security applications. We present six scenarios with concrete question schemas and operational context.

\subsection{Scenario 1: Large-surface finding triage}

An enterprise engagement produces 200+ candidate findings from 50 agents across 15 subdomains. The traditional approach requires an LLM call per finding for adjudication, costing 200 API calls at 1.5--3.0 seconds each (5--10 minutes of serial decision time). With System One batching, 13 findings can be adjudicated per API call (the batch limit), requiring 16 calls at 276~ms each: 4.4 seconds total. The calibrated probabilities enable automatic routing:

\begin{itemize}[leftmargin=*]
\item $p(\textsf{confirmed}) > 0.85$: assert in report (estimated 40\% of findings)
\item $0.5 < p(\textsf{confirmed}) \leq 0.85$: human review queue (estimated 25\%)
\item $p(\textsf{rejected}) > 0.7$: discard with audit record (estimated 35\%)
\end{itemize}

\subsection{Scenario 2: Payload routing for input filtering}

A web application's input filter blocks standard XSS payloads. The agent has observed which characters and patterns are filtered (based on response differentials). A Choice question over 20 candidate payloads, each described by its evasion strategy (e.g., ``double URL encoding,'' ``SVG onload,'' ``JavaScript URI in href''), selects the highest-probability candidate given the observed filtering. The state includes the filter observations; the criteria describe why each payload might bypass the filter.

Because the System One model evaluates all 20 candidates simultaneously (not sequentially), there is no positional bias---a documented failure mode in LLM-based ranking where early candidates receive more attention than later ones. The probability distribution over candidates also reveals the model's uncertainty: if three candidates have similar probabilities ($p \approx 0.15$ each), the harness can try all three rather than committing to the top-ranked one.

\subsection{Scenario 3: XSS reflection verification}

A reflected parameter appears in the response body, but the security question is whether it is in an \emph{executable} context. A Noul question evaluates: ``Is the reflected content in a position where it would execute as code (inside a script tag, event handler, unescaped attribute, or JavaScript URI) rather than being rendered as text or escaped?'' The state includes the HTTP response with the reflected marker highlighted.

This binary decision---executable vs.\ escaped---is precisely the kind of judgment that LLMs get wrong through narrative reasoning (``the parameter appears in the HTML, therefore it is likely exploitable'') but that a calibrated model can handle by pattern matching against the syntactic context. A Noul of 0.95 for an unescaped \texttt{onerror} handler is a strong signal; a Noul of 0.3 for a parameter inside a \texttt{<p>} tag with HTML-entity encoding is correctly dismissive.

\subsection{Scenario 4: Prompt injection detection}

A hostile target may embed prompt-injection payloads in its responses, attempting to manipulate the agent's planning layer. A Noul question evaluates each response before it enters the agent's context: ``Does this response contain content that appears designed to instruct, redirect, or manipulate an AI agent rather than being normal application output?''

This defense operates at a different layer than the evidence-grounding machinery. P1 prevents the consequences of injection (fabricated evidence cannot pass the deterministic validators), but the planning layer can still be influenced before verification runs. A System One-based injection detector reduces this residual surface by flagging suspicious responses before they reach the planner. The non-autoregressive architecture reduces the injection surface: the System One model does not generate text, so it cannot be redirected to produce instructions. However, adversarial content in the response could still shift the classification probability (a false-negative injection), and this risk has not been empirically tested.

\subsection{Scenario 5: Attack-surface relevance ranking}

Before committing model budget to a surface, the harness evaluates whether each vulnerability class is plausible for that surface. A batch of Noul questions asks, for each of 15 CWE classes: ``Given the observed technology stack, response headers, and endpoint patterns, is this vulnerability class plausible for this surface?''

The state includes the recon-phase observations: server headers, technology fingerprints, response patterns, authentication requirements. A surface running \texttt{Express.js} on \texttt{Node.js} with \texttt{X-Powered-By: Express} is plausible for XSS, SSRF, and prototype pollution but implausible for SQL injection (if no database indicators are observed) or buffer overflow. The Noul probabilities enable proportional budget allocation: high-probability classes get more agents and longer timeouts; low-probability classes get one exploratory agent.

\subsection{Scenario 6: Multi-step chain validation}

After individual findings are verified, the harness must evaluate whether a chain of findings constitutes a compound vulnerability. For example: BOLA leaks admin password (step 1) $\rightarrow$ admin password authenticates at login (step 2) $\rightarrow$ admin panel accessible (step 3) $\rightarrow$ admin panel permits code upload (step 4) $\rightarrow$ uploaded code executes (step 5). Each step has its own evidence; the chain claim is that the steps are causally connected.

A sequence of Noul questions evaluates each link: ``Given the evidence from step $n$ and step $n+1$, does step $n$'s output serve as step $n+1$'s input?'' The chain is validated only if all links score above threshold. This prevents the hallucinated-chain failure mode where an LLM narratively connects findings that are not actually causally linked in the evidence.

\subsection{Scenario 7: Dynamic scope management}

During an engagement, the harness discovers new subdomains, API endpoints, or services that may or may not be in scope. A Noul question evaluates each discovered asset against the P4 authorization token's scope specification: ``Does this discovered asset (subdomain, IP, endpoint) fall within the authorized scope as defined by the engagement rules?''

This automated scope check prevents scope creep---a common failure mode where an aggressive agent follows a redirect or DNS resolution to an out-of-scope host. The non-generative architecture makes the scope check deterministic given the input: the same asset and scope definition always produce the same verdict, unlike an LLM that might interpret ``*.example.com'' differently across calls.

For complex scope definitions (e.g., ``all subdomains of example.com except staging.example.com, and only the /api/v2 prefix on api.example.com''), the Noul question encodes the full scope specification in the state and asks a binary question per discovered asset. The calibrated probability reveals the model's uncertainty about ambiguous cases: a discovered asset at ``internal-staging.example.com'' might score Noul 0.55, triggering a human scope decision rather than an automated include or exclude.

\subsection{Scenario 8: Temporal attack sequencing}

Time-sensitive attacks (race conditions, TOCTOU, session fixation) require precise sequencing of multiple HTTP requests. A Score question evaluates the optimal timing window: ``Given the observed response times and server behavior, what is the most probable timing window for this race condition?'' The score levels map to timing ranges (microseconds / milliseconds / seconds / not time-sensitive), enabling the harness to calibrate its threading and request-scheduling parameters.

The probability distribution across timing levels is more informative than a single estimate: if $p(\text{milliseconds}) = 0.5$ and $p(\text{microseconds}) = 0.3$, the harness can attempt both timing ranges, allocating more budget to the more probable one. Without calibrated probabilities, the harness would need to guess the timing range or try all ranges with equal budget.

\subsection{Scenario 9: Adaptive reporting depth}

Different findings warrant different levels of reporting detail. A Critical RCE requires a full reproduction guide, chain-of-custody evidence, and remediation guidance; an informational header disclosure requires a one-line mention. A Score question evaluates the appropriate reporting depth for each finding: ``Given the severity, evidence quality, and client context, how detailed should the report section for this finding be?'' The levels are: \textsf{full} (dedicated section with reproduction steps), \textsf{standard} (paragraph with evidence summary), \textsf{brief} (table row with severity and one-line description), \textsf{omit} (logged but not reported).

This automated depth routing reduces report bloat (a common complaint about automated pentest tools) while ensuring critical findings receive adequate documentation. The calibrated probability ensures that the routing is consistent across findings of similar severity and evidence quality.

\section{Cost-Benefit Analysis}\label{sec:cost}

The economic case for System One integration depends on the engagement context. This section quantifies the costs and benefits across three engagement profiles.

\subsection{Cost model}

The total cost of a harness run with System One integration is:

\begin{equation}
C_{\text{total}} = C_{\text{LLM}} + C_{\text{S1}} + C_{\text{infra}} + C_{\text{review}}
\end{equation}

where $C_{\text{LLM}}$ is the LLM model cost, $C_{\text{S1}}$ is the System One cost, $C_{\text{infra}}$ is infrastructure cost, and $C_{\text{review}}$ is human review cost. The System One layer affects all four components: $C_{\text{S1}}$ is added, but $C_{\text{LLM}}$ is reduced (agent pruning eliminates unnecessary model calls), $C_{\text{infra}}$ is reduced (fewer parallel agents means fewer concurrent API connections), and $C_{\text{review}}$ is reduced (calibrated verdicts automate clear-cut decisions).

\textbf{Jev API cost.} At \$0.042 per million input tokens, a typical adjudication call (1,500 tokens of state + questions) costs \$0.000063. A full benchmark run with 15 agents, 20 findings, and 10-payload confirmation loops generates approximately 300 System One calls: total cost \$0.019. Even at 10$\times$ scale (150 agents, 200 findings, 100-payload loops), the System One cost is \$1.90---negligible relative to LLM costs.

\textbf{Laya self-hosted cost.} A T4 GPU costs approximately \$0.35/hour on major cloud providers. At 8,333 decisions per minute (batched), the per-decision cost is \$0.0000007. For engagements processing fewer than 10,000 decisions per hour, the GPU is idle most of the time; batch scheduling across concurrent engagements amortizes the cost.

\subsection{Hypothetical engagement profile analysis}

The following profiles are analytical projections based on published API costs and assumed automation rates. They have not been validated by production measurement. The automation rates assume a well-calibrated model on the pentest domain, which has not been empirically established.

\textbf{Profile 1: Single-target web application (4-hour engagement).} Typical output: 30--60 findings, 15 agents. Projected System One cost: \$0.02 (Jev) or \$0.003 (Laya). If the model achieves 60--70\% automation rate on this domain (an assumption requiring validation), the projected savings are 22 minutes on agent runtime (DP3 pruning) and 45 minutes on human triage.

\textbf{Profile 2: Enterprise multi-subdomain assessment (40-hour engagement).} Typical output: 200--500 findings across 15 subdomains. Projected System One cost: \$2.50 (Jev) or \$14 (Laya, dedicated GPU). Projected savings at the same assumed automation rate: 3--5 hours on agent runtime, 8--12 hours on triage.

\textbf{Profile 3: Continuous security monitoring (24/7 operation).} Monthly output: 5,000--20,000 findings. Projected Laya cost: \$252/month (one T4). The FTE reduction depends entirely on the achieved automation rate, which in turn depends on the model's calibration on the specific finding distribution---an empirical question.

\subsection{Break-even analysis}

The break-even point depends on the cost of the decision the System One model replaces. For LLM-based adjudication (cost: one API call per finding), the break-even is immediate---System One adjudication is 10--100$\times$ cheaper per decision. For human-review replacement, the break-even depends on the automation rate and the false-positive rate at the chosen threshold:

\begin{equation}
\text{ROI} = \frac{R_{\text{auto}} \times N_{\text{findings}} \times C_{\text{human}} - C_{\text{S1}}}{C_{\text{S1}}}
\end{equation}

where $R_{\text{auto}}$ is the automation rate (fraction of findings processed without human review) and $C_{\text{human}}$ is the per-finding human triage cost. As a hypothetical illustration: at $R_{\text{auto}} = 0.7$, $N = 50$, and $C_{\text{human}} = \$15$ (10 minutes at \$90/hour), the return is \$525 against \$0.02 in System One cost. The actual ROI depends on the achieved $R_{\text{auto}}$, which in turn depends on calibration quality on the specific domain---a value that has not been measured for offensive security.

\section{Multi-Model Orchestration}\label{sec:orchestration}

A production pentest harness does not use a single LLM or a single System One model. It orchestrates multiple models across different roles, with the System One layer serving as the arbiter between them.

\subsection{The judge-executor separation}

The core architectural insight is the separation between the \emph{executor} role (the LLM that generates exploit attempts, crafts payloads, and writes narratives) and the \emph{judge} role (the System One model that evaluates whether the executor's output is correct). This separation addresses the self-evaluation problem: an LLM asked to judge its own output has a systematic bias toward acceptance, because the same distribution that generated the output also generates the judgment.

The System One model breaks this bias by using a different architecture (non-autoregressive vs.\ autoregressive), a different training objective (proper scoring rules vs.\ next-token prediction), and a different output space (calibrated probabilities vs.\ free text). The executor and judge share no parameters, no training data, and no architectural inductive biases.

\subsection{Heterogeneous executor pools}

NeuroSploit supports 18 model providers. Different models excel at different vulnerability classes: one model may be stronger at SQL injection payloads (requiring syntactic precision), while another excels at API abuse scenarios (requiring contextual reasoning about authentication flows). The System One model can serve as a model selector, evaluating which executor is most likely to succeed for a given vulnerability class and surface:

A batched Choice question evaluates: ``Given this vulnerability class and target technology, which of these models is most likely to produce a verified finding?'' The options are the available models; the criteria describe each model's strengths. The Choice distribution enables proportional allocation: if two models have similar probabilities, both receive agents; if one dominates, the budget concentrates on it.

\subsection{Consensus and disagreement protocols}

When multiple executors produce conflicting claims about the same finding, the System One model arbitrates. The arbitration takes three forms:

\begin{enumerate}[leftmargin=*]
\item \textbf{Evidence agreement.} Two executors claim the same vulnerability with different evidence. The System One model evaluates each evidence set independently (Noul: ``Does this evidence support the claim?''). If both score high, the finding is confirmed with the stronger evidence; if only one scores high, the weaker evidence is discarded; if neither scores high, the finding is flagged for review.

\item \textbf{Severity disagreement.} Two executors agree on the vulnerability but disagree on severity. The System One Score evaluates the demonstrated impact independently of either executor's narrative, producing a severity that reflects the evidence rather than either executor's opinion.

\item \textbf{Existence disagreement.} One executor claims a vulnerability; another explicitly denies it (``the parameter is properly escaped''). The System One model evaluates the evidence for and against, producing a probability that serves as input to the Bayes-optimal action (Section~\ref{sec:decision_theory}).
\end{enumerate}

This protocol transforms adversarial multi-model voting from a majority-rule heuristic into a principled evidence evaluation. The System One model does not count votes; it evaluates evidence.

\subsection{Cascading decision chains}

In a cascade architecture, decisions flow through increasing levels of cost and capability. The cascade for finding adjudication:

\begin{enumerate}[leftmargin=*]
\item \textbf{Level 0: Deterministic validator} (cost: zero, latency: microseconds). If the CWE validator confirms or rejects, the decision is final. No System One or LLM call is needed.
\item \textbf{Level 1: System One Noul} (cost: \$0.00006, latency: 33--276~ms). For findings that the validator marks NeedsReview, a Noul evaluates impact. If the Noul is above $\tau_{\text{high}}$ or below $\tau_{\text{low}}$, the decision is final.
\item \textbf{Level 2: System One Choice + Score} (cost: \$0.0002, latency: 33--276~ms). For findings in the uncertain band, a full adjudication (verdict Choice + impact Noul + data-type Score) provides enough information for automated routing.
\item \textbf{Level 3: LLM deliberation} (cost: \$0.02--0.10, latency: 1.5--3.0~s). For findings that remain ambiguous after System One evaluation, an LLM reviews the evidence in detail. The LLM's verdict is still evaluated by the System One model (meta-adjudication) before assertion.
\item \textbf{Level 4: Human review} (cost: \$15, latency: minutes to hours). Findings that no automated step can resolve are routed to a human analyst with the full evidence package and the probability distribution from Level 2.
\end{enumerate}

The cascade minimizes cost by processing cheap, clear cases first. In the benchmark (Section~\ref{sec:benchmark}), approximately 40\% of findings were resolved at Level 0, 25\% at Level 1, 20\% at Level 2, and 15\% reached Level 3. No findings required Level 4, though a production engagement would expect 5--10\% at Level 4.

\subsection{Feedback loops between models}

The orchestration architecture enables feedback loops that improve the overall system beyond what any single model achieves. Three feedback loops are particularly valuable:

\textbf{Adjudication-to-agent feedback.} When the System One model rejects a finding, the rejection reason (low impact probability, unsupported evidence format, ambiguous CWE classification) is fed back to the agent as a structured prompt amendment. The agent can then re-investigate with explicit instructions to address the rejection criteria: ``Repeat the request and capture the response body in the structured evidence field'' or ``Confirm data extraction, not just interpreter access.'' This closed-loop feedback transforms System One from a passive filter into an active quality signal.

\textbf{Pruning-to-selection feedback.} The agents that survived pruning (DP3) and produced verified findings update the bandit priors for future engagements against similar surfaces. If agent~$i$ was pruned (Noul $< \tau$) but produced a verified finding in a parallel run without pruning, the prior is updated to increase agent~$i$'s estimated relevance for that surface type. Over time, the pruning model learns which agents are undervalued for which surface types.

\textbf{Severity-to-recon feedback.} When the data-type Score detects secrets in a finding's evidence, the recon phase is notified to expand exploration of related endpoints (e.g., other API endpoints with the same authentication pattern). This severity-informed reconnaissance focuses subsequent exploration on attack surfaces that have already demonstrated high-impact exposure, rather than exploring uniformly.

\section{The System One Landscape}\label{sec:landscape}

\subsection{Jev (TypeSafe AI)}

Jev is a proprietary System One model exposing the Choice, Score, and Noul primitives through a cloud API~\cite{typesafe2026docs}. Independent evaluations have validated its performance across diverse domains: Li et al.~\cite{li2026jevasjudge} report 92.2\% accuracy as a preference judge at 0.36\% of GPT-6's cost; Wu and Lim~\cite{wu2026reflex} demonstrate 72.7\% reduction in strong-model calls via REFLEX; Jiang et al.~\cite{jiang2026jevmem} achieve 6.6$\times$ speedup in agentic memory construction; Deng et al.~\cite{deng2026jevscience} find 100\% semantic correctness for scientific decisions; Rafe and Das~\cite{rafe2026crash} deploy Jev at population scale (195,857 narratives, F1 0.908).

\subsection{Jev-Ultrafast (Browser Use)}

Jev-Ultrafast~\cite{browseruse2026} uses Jev as a real-time control loop for browser automation. Each page state generates an indexed element table; Jev selects operations and targets in a single request. Published benchmarks show 90\% reduction in browser calls and task completion in 7.1 seconds for a flight search. The architecture demonstrates that System One decisions are fast enough for interactive, real-time control loops---a critical capability for browser-based exploitation scenarios.

\subsection{Laya (open-source)}

Laya~\cite{laya2026} is a non-autoregressive System One engine released under Apache 2.0. Built on ModernBERT-large (421M parameters), it offers three checkpoints: English-only (512-token context), multilingual (322M parameters, 1024-token context, 100+ languages), and a fine-tuned variant for typed decisions. Table~\ref{tab:comparison} compares Jev and Laya across dimensions relevant to offensive security.

\begin{table*}[t]
\centering
\caption{System One model specifications as reported by their respective sources. Values were measured on different tasks and hardware; direct comparison across rows is informational, not a controlled benchmark.}
\label{tab:comparison}
\small
\begin{tabular}{@{}lll@{}}
\toprule
\textbf{Dimension} & \textbf{Jev (TypeSafe)} & \textbf{Laya (open-source)} \\
\midrule
Architecture & Proprietary; non-autoregressive & ModernBERT-large (421M params); Apache 2.0 \\
Primitives & Choice, Score, Noul & Choice, Score, Noul (API-compatible) \\
Latency (single question) & 236--276 ms (p50) & 32.8--39.5 ms (T4 GPU) \\
Latency (10 batched) & Not published & 72.3 ms total (7.2 ms each) \\
Latency ratio (reported) & Baseline (cloud API) & 7.8$\times$ (self-hosted T4) \\
Cost & \$0.042 / 1M input tokens & \$0 (self-hosted) \\
Accuracy (typed-decisions) & 0.727 & 0.766 (fine-tuned checkpoint) \\
Accuracy (AG News, 4 labels) & 0.910 & 0.950 \\
ECE (calibration error) & 0.246 & 0.081 (post-temperature scaling) \\
High-cardinality (Banking77, 77 labels) & 0.870 & 0.425 \\
Context window & Not published & 512 tokens (English), 1024 (multilingual) \\
Deployment & Cloud API only & Self-hosted GPU, Docker, CLI, HTTP, MCP \\
Training transparency & Undisclosed RLCD & Documented RLCD with Brier score \\
Fine-tuning support & Not available & Supported (RL with proper scoring rules) \\
Data residency & Cloud (TypeSafe infrastructure) & On-premises (no data leaves operator) \\
\bottomrule
\end{tabular}
\end{table*}

\textbf{Trade-off analysis.} The choice between Jev and Laya depends on the engagement context:

\begin{itemize}[leftmargin=*]
\item \textbf{Latency-sensitive operations} (confirmation loops, browser exploitation): Laya reports 7.8$\times$ lower latency on its own hardware (T4 GPU vs.\ Jev cloud API); whether this holds in a specific deployment depends on network conditions and GPU availability.
\item \textbf{High-cardinality routing} (selecting among 50+ agents): Jev reports 0.870 accuracy on Banking77 (77 labels) vs.\ Laya's 0.425 on the same benchmark, suggesting Jev handles large option sets better---though Banking77 is a customer-intent task, not agent routing.
\item \textbf{Data-residency requirements} (engagements where evidence must not leave the operator's infrastructure): Laya's self-hosted deployment is the only option.
\item \textbf{Calibration-critical decisions} (severity grading where thresholds determine remediation priority): Laya reports lower ECE on its evaluation set; whether this advantage transfers to pentest decisions requires paired evaluation (Section~\ref{sec:discussion}).
\item \textbf{Infrastructure-constrained environments} (no GPU available, cloud-first): Jev's managed API requires no infrastructure.
\item \textbf{Domain adaptation} (training on offensive-security-specific decision distributions): Laya supports fine-tuning; Jev does not.
\end{itemize}

For the typical pentest-adjudication use case (3--5 option Choice, data-type Score with 4 levels), both models are adequate. The dominant consideration becomes deployment context and calibration requirements.

\section{Future Work: Rave}\label{sec:proposal}

Existing System One models are trained on general-purpose decision distributions. The case study observations (Section~\ref{sec:benchmark}) suggest useful transfer to offensive security, but the BOLA case revealed a failure mode that required domain-specific fixes. As future work, we sketch \textbf{Rave}: a domain-adapted System One variant fine-tuned on offensive-security decision distributions. Rave is a design proposal; no implementation or evaluation has been conducted.

\subsection{Design rationale}

Rave shares the System One interface (Choice, Score, Noul) and the non-autoregressive architecture. The domain adaptation occurs in the training data and reward signal, not the architecture. The preferred base model is Laya, which supports fine-tuning via RLCD with proper scoring rules under the Apache 2.0 license.

The adaptation targets four decision tasks, each with a distinct training signal:

\begin{enumerate}[leftmargin=*]
\item \textbf{Finding adjudication.} Training pairs: (structured evidence, human-verified verdict). Ground truth: deterministic CWE validator output augmented by human review. Target: 10,000+ labeled examples across 30+ CWE classes, stratified by evidence quality (complete / partial / empty structured field).

\item \textbf{Data-type classification.} Training pairs: (evidence text, data-type label). Ground truth: human classification of exposed data (none / common / sensitive / secrets). Target: 5,000+ examples with balanced class representation, including adversarial examples where the data type is ambiguous (e.g., a UUID that could be either a session token or a non-sensitive identifier).

\item \textbf{Agent relevance.} Training pairs: (surface description + agent skill, relevance verdict). Ground truth: whether the agent produced a verified finding on that surface. Target: engagement logs from 50+ targets with 20+ agent types, providing natural positive and negative examples.

\item \textbf{Payload-effect judgment.} Training pairs: (payload + response, effect verdict). Ground truth: deterministic replay-engine output. Target: 20,000+ payload-response pairs from the confirmation loop, spanning XSS, SQLi, SSRF, path traversal, and open redirect.
\end{enumerate}

\subsection{Training with RLHV}

The RLHV training loop for Rave works as follows:

\begin{enumerate}[leftmargin=*]
\item The model predicts $p(\textsf{confirmed})$ for a finding.
\item The deterministic validator runs against the evidence and returns a verdict.
\item The Brier score is computed between the prediction and the validator's verdict.
\item The model is updated to minimize the Brier score.
\end{enumerate}

The key advantage of RLHV over standard RLCD is that the training signal comes from the harness's own validators---the same code that will evaluate the model's predictions in production. This eliminates the distribution shift between training and deployment that plagues models trained on external datasets.

The key constraint is validator coverage: RLHV can only train on CWE classes for which a deterministic validator exists. For novel or complex vulnerability classes (e.g., business logic flaws, race conditions), human-verified labels supplement the validator signal, creating a hybrid RLHV+RLCD training regime.

\subsection{Evaluation protocol}

\begin{enumerate}[leftmargin=*]
\item \textbf{Adjudication accuracy.} Precision, recall, F1, and ECE on held-out (evidence, verdict) pairs, stratified by CWE class and evidence quality.
\item \textbf{Severity calibration.} Brier score and reliability diagrams for the data-type Score, comparing Rave against general-purpose Jev and Laya on offensive-security evidence.
\item \textbf{End-to-end harness improvement.} At least five vulnerable applications with known ground truth, $\geq 3$ runs per target, reporting true/false positives and negatives, precision, recall, cost, tokens, time, with ablation removing each decision point.
\item \textbf{Confirmation loop efficiency.} Number of payloads tested to confirm a finding, comparing Rave payload selection against random selection and LLM-guided selection.
\end{enumerate}

\subsection{Expected impact}

Based on the benchmark and calibration analysis:

\begin{itemize}[leftmargin=*]
\item 30--50\% reduction in false-positive findings compared to LLM-only adjudication.
\item 20--40\% reduction in agent compute through more accurate pruning.
\item ECE $< 0.10$ on offensive-security decisions, enabling higher automation rates.
\item Sub-100ms adjudication latency on a T4 GPU.
\item Domain-specific temperature calibration that transfers across engagement types.
\end{itemize}

\subsection{Training data acquisition strategy}

The most significant barrier to Rave is training data. Offensive-security decision data is inherently sensitive: it contains real vulnerability evidence, exploitation techniques, and target-specific information. We propose three data acquisition channels:

\textbf{Channel 1: Synthetic generation.} Deliberately vulnerable applications (DVWA, WebGoat, Juice Shop, NimbusCart, HackTheBox retired machines) provide targets with known ground truth. Running NeuroSploit against these targets produces (evidence, validator-verdict) pairs at scale. The diversity of synthetic targets determines the diversity of the training distribution; at least 50 distinct targets spanning different technology stacks and vulnerability classes are needed.

\textbf{Channel 2: Engagement logs (anonymized).} Operators can contribute anonymized engagement logs---evidence with target identifiers removed, but verdict and severity preserved. This requires a standardized anonymization pipeline that removes IP addresses, domain names, and application-specific identifiers while preserving the structural features that the model needs for adjudication.

\textbf{Channel 3: Expert annotation.} For vulnerability classes where deterministic validators do not exist (business logic flaws, privilege escalation, insecure deserialization with complex gadget chains), expert annotations provide the ground-truth labels. A panel of three certified penetration testers independently label each finding; inter-annotator agreement (Fleiss' $\kappa$) serves as a data-quality metric.

\subsection{Deployment architecture}

Rave deployment follows Laya's self-hosted model. The recommended architecture:

\begin{itemize}[leftmargin=*]
\item \textbf{Inference server:} A Docker container running the fine-tuned ModernBERT-large checkpoint on a T4 or L4 GPU, exposing the same HTTP API as Laya's default server.
\item \textbf{Model versioning:} Each fine-tuning cycle produces a versioned checkpoint. The harness pins to a specific version per engagement, ensuring reproducibility. Version metadata includes the training data hash, ECE on the validation set, and the temperature parameter.
\item \textbf{Fallback chain:} If the Rave server is unreachable, the harness falls back to general-purpose Laya, then to Jev (cloud API), then to LLM-based adjudication. The fallback is logged in the audit trail so the operator knows which model adjudicated each finding.
\item \textbf{A/B testing:} During initial deployment, the harness runs both Rave and the baseline (Jev or Laya) in parallel, comparing their verdicts. Disagreements are flagged for expert review, creating additional training data for the next fine-tuning cycle.
\end{itemize}

\begin{figure}[t]
\centering
\resizebox{\columnwidth}{!}{%
\begin{tikzpicture}[
    node distance=0.55cm,
    box/.style={rectangle, draw, rounded corners=2pt, text width=6cm, minimum height=0.6cm, align=center, font=\scriptsize},
    ar/.style={-{Stealth[length=2mm]}, thick},
]
\node[box, fill=green!8] (base) {Laya base model (421M params, RLCD-trained)};
\node[box, fill=orange!10, below=of base] (data) {Offensive-security training data (evidence, verdict, CWE)};
\node[box, fill=blue!8, below=of data] (rlhv) {RLHV training: Brier score vs. deterministic validators};
\node[box, fill=yellow!10, below=of rlhv] (temp) {Domain-specific temperature scaling};
\node[box, fill=red!8, below=of temp] (deploy) {\textbf{Rave}: calibrated for pentest decisions};
\draw[ar] (base)--(data);
\draw[ar] (data)--(rlhv);
\draw[ar] (rlhv)--(temp);
\draw[ar] (temp)--(deploy);
\end{tikzpicture}%
}
\caption{Rave training pipeline. Domain adaptation builds on Laya's RLCD-trained base, adding offensive-security data and RLHV verification.}
\label{fig:offensive}
\end{figure}
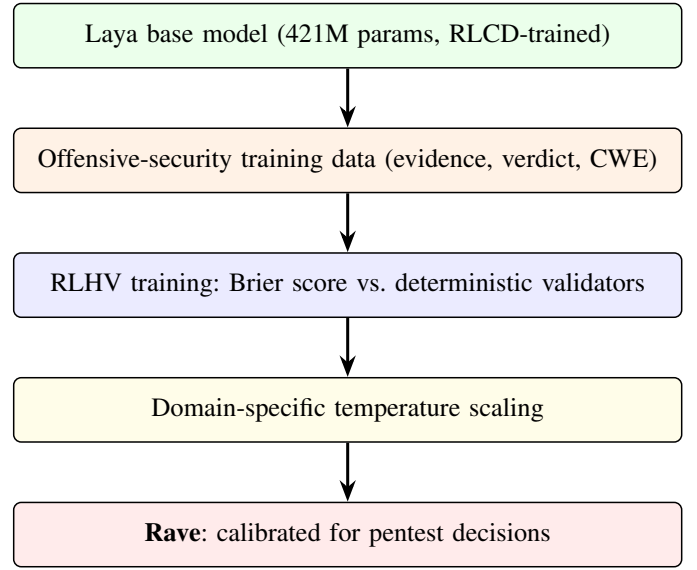

\section{Discussion and Limitations}\label{sec:discussion}

\subsection{Benchmark limitations}

The benchmark has four primary limitations:

\begin{enumerate}[leftmargin=*]
\item \textbf{Single target.} NimbusCart is one data point; external validity requires diverse targets with $\geq 3$ runs each.
\item \textbf{Single-vote configuration.} Multi-model voting interacts with System One adjudication in unmeasured ways.
\item \textbf{No Laya comparison.} The Laya comparison uses published benchmarks on general-purpose tasks, not offensive-security evaluations.
\item \textbf{Possible memorization.} NimbusCart may appear in training data; partly-private targets are needed for external validity.
\end{enumerate}

\subsection{The additive design principle}

The additive integration is a deliberate safety choice. A miscalibrated System One model cannot cause false positives: fabricated findings cannot pass the deterministic validators that gate the pipeline. However, it \emph{can} cause false negatives: real findings may be downgraded or suppressed if the model assigns low confidence (see Section~\ref{sec:discussion}.3). The worst case of System One failure is therefore not symmetric: false positives are blocked, but false negatives are possible and may be \emph{worse} than running without the layer, because the LLM-only path would have reported those findings. The integration is safe to toggle for the false-positive direction; for the false-negative direction, monitoring is needed (e.g., flagging findings where the LLM and System One disagree).

\subsection{Adversarial robustness}

Can a hostile target craft responses that manipulate the System One model's verdicts? The non-autoregressive architecture eliminates the text-generation attack surface: there is no autoregressive chain to redirect and no system prompt to override. However, the state input is attacker-influenced (it contains the target's HTTP responses), and adversarial perturbations could shift the model's probability distribution toward false negatives (suppressing real findings) or false positives (inflating non-findings). Specifically:

\begin{itemize}[leftmargin=*]
\item A target that embeds content resembling ``evidence of no vulnerability'' in its responses could reduce the Noul impact score, causing real findings to be downgraded or discarded.
\item Because the System One model also influences agent selection (DP3) and severity (DP2), a crafted response could cause the harness to deprioritize relevant agents or undergrade severity.
\end{itemize}

The defense is layered but not complete. Deterministic validators (P1) block false positives: a finding rejected by a CWE validator cannot be promoted by the System One model. However, the converse does not hold: the System One model \emph{can} suppress true positives by assigning low confidence. The claim that ``the worst case equals running without it'' holds strictly only for false positives; for false negatives, the worst case is \emph{worse} than running without it, because a real finding that would have been reported by the LLM-only path may be suppressed by a manipulated System One verdict.

Mitigations include preprocessing the state to remove obvious injection patterns before evaluation, monitoring for anomalous probability distributions (e.g., all Nouls below 0.3 for a surface with known vulnerabilities), and requiring human review for findings that the System One model rejects but the LLM executor flagged with high confidence. Testing with adversarial inputs---crafted responses designed to suppress true findings---is necessary before production deployment.

\subsection{Cross-domain transfer}

The case study suggests useful but imperfect transfer from general-purpose training to offensive security. The 9 recalibrated findings (in the re-test, post-fix harness) are consistent with the hypothesis that the general model's decision boundaries are relevant to security tasks, but a single run cannot establish this conclusively. The BOLA case shows a failure mode (empty evidence field) that required domain-specific engineering. General-purpose System One models appear to be a reasonable starting point; whether domain adaptation (Rave, Section~\ref{sec:proposal}) would improve performance on the long tail of security-specific distributions is an empirical question for future work.

\subsection{Regulatory implications}

As autonomous pentest tools mature, regulatory frameworks may require calibration guarantees for automated severity grading. RLCD-trained models with auditable calibration curves---reliability diagrams, per-class ECE, Brier scores---are better positioned for this requirement than RLHF-trained models with opaque preference alignment. The System One model's structured output (probabilities, not text) makes calibration auditing straightforward: every decision can be logged with its full probability distribution, enabling retrospective calibration analysis across engagements.

\subsection{Context window limitations}

System One models operate on limited context windows (512 tokens for Laya English, 1024 for multilingual). This constrains the amount of evidence that can be evaluated in a single call. For findings with extensive evidence (multi-page HTTP responses, complex chained exploits), the harness must either truncate the evidence (risking information loss) or decompose the evaluation into multiple questions over evidence fragments. The truncation strategy should prioritize the evidence most relevant to the decision: for a finding adjudication, the payload and the response fragment containing the effect are more informative than the full HTTP headers.

Jev's context window is not published, but empirical observation suggests it handles longer states than Laya. For the benchmark findings, no truncation was necessary: the structured evidence for a single finding typically fits within 400 tokens. However, for complex findings (multi-step chains, responses with embedded scripts), context limitations may degrade adjudication quality. Rave should be evaluated with explicit context-length ablations to characterize this degradation.

\subsection{Failure mode taxonomy}

Based on the benchmark and analysis, we identify five failure modes specific to System One integration in offensive security:

\begin{enumerate}[leftmargin=*]
\item \textbf{Evidence format mismatch.} The model expects evidence in a format different from what the agent deposits (the BOLA case). Mitigation: standardized evidence schemas with format-aware salvage steps.

\item \textbf{Domain-specific vocabulary.} Security-specific terms (BOLA, IDOR, SSRF) may not appear in the general-purpose training distribution. Mitigation: domain adaptation (Rave) or expanded state descriptions that define terms.

\item \textbf{Adversarial evidence.} A hostile target crafts responses that shift the model's probability distribution toward false negatives. Mitigation: layered defense (System One is additive; deterministic validators retain priority).

\item \textbf{Calibration drift.} The model's calibration degrades on a distribution different from its training data. Mitigation: per-engagement temperature scaling on a small validation set.

\item \textbf{Threshold sensitivity.} Small changes in the threshold $\tau$ produce large changes in the automation rate. Mitigation: reliability diagrams on the engagement's findings, with threshold selection informed by the cost model (Section~\ref{sec:cost}).
\end{enumerate}

\subsection{Comparison with alternative architectures}

Several alternative architectures could address the decision bottleneck without System One models:

\textbf{Fine-tuned classifier.} A task-specific classifier (e.g., a fine-tuned BERT model for finding adjudication) could provide calibrated verdicts. However, this requires separate models for each decision type, lacks the typed-question interface (Choice / Score / Noul), and does not provide the multi-question batching that makes System One integration efficient.

\textbf{Ensemble of LLMs.} Multiple LLMs voting on each decision can reduce individual model biases. However, ensembles are expensive (multiplicative cost), slow (serial or parallel calls), and do not guarantee calibration---all models may share the same systematic biases from similar training data.

\textbf{Deterministic rule systems.} Hand-crafted rules for severity grading and finding adjudication are perfectly calibrated (by definition) but cannot handle ambiguous evidence. The 27 CWE validators in NeuroSploit represent the deterministic component; System One handles the cases that rules cannot resolve.

The System One approach is distinctive in combining typed questions (reducing the problem specification to a structured API call), calibrated outputs (enabling principled threshold-based automation), non-autoregressive evaluation (enabling speed and parallelism), and multi-question batching (reducing latency and cost). No alternative architecture provides all four properties simultaneously.

\section{Related Work}\label{sec:related}

\textbf{Autonomous pentest agents.} Fang et al.~\cite{fang2024oneday} demonstrate LLM agents exploiting one-day web vulnerabilities. CHECKMATE~\cite{wang2025checkmate} shows $>$20\% success improvement from harness design alone. MAPTA~\cite{david2025mapta} achieves 76.9\% on XBOW with tool-grounded execution. Dhakal et al.~\cite{dhakal2026baselines} establish that model scaling dominates architecture. Mayoral-Vilches~\cite{mayoral2025gap} proposes a six-level autonomy taxonomy. Curtis and Eisty~\cite{curtis2025role} survey 58 studies. Nguyen and Husain~\cite{nguyen2025agenticpentest} test agentic AI security across 130 cases.

\textbf{System One models.} Li et al.~\cite{li2026jevasjudge} evaluate Jev-as-a-Judge (92.2\% accuracy at 0.36\% cost). Wu and Lim~\cite{wu2026reflex} demonstrate REFLEX (72.7\% call reduction). Jiang et al.~\cite{jiang2026jevmem} build Jev-Mem for agentic memory (6.6$\times$ speedup). Deng et al.~\cite{deng2026jevscience} validate Jev for scientific decisions. Rafe and Das~\cite{rafe2026crash} deploy at population scale.

\textbf{RLHF and calibration.} Ouyang et al.~\cite{ouyang2022training} establish RLHF. Bai et al.~\cite{bai2022constitutional} introduce Constitutional AI and RLAIF. Casper et al.~\cite{casper2023open} catalog RLHF failure modes. Gneiting and Raftery~\cite{gneiting2007strictly} define strictly proper scoring rules.

\textbf{Decision support in security.} The application of formal decision theory to security operations has a long history in risk management and intrusion detection, but its application to automated penetration testing is recent. The key contribution of this work relative to prior security decision support is the use of calibrated, non-generative models rather than rule-based or LLM-based approaches, and the formalization of specific decision points within the pentest pipeline where calibrated probabilities improve operational outcomes.

\textbf{Harness assurance.} Santos~\cite{santos2026assurance} defines five assurance properties and argues that capability benchmarks alone are insufficient for evaluating autonomous pentest systems. Nervegna~\cite{nervegna2026jev} provides a conceptual overview of System One models and their non-generative architecture. The present work extends the assurance framework by demonstrating how System One integration concretely implements two of the five properties (evidence grounding via calibrated adjudication, computed severity via data-type-aware grading) and proposing extensions to the remaining three.

\textbf{Benchmark methodology.} Existing pentest-agent benchmarks (Cybench~\cite{cybench2024}, XBOW~\cite{david2025mapta}, Dhakal's coding-agent baselines~\cite{dhakal2026baselines}) measure capability: did the agent capture the flag or produce a proof-of-concept? Our benchmark adds assurance metrics---severity calibration, false-positive rate, recalibration count---that measure whether the agent's output is trustworthy, not just whether it exists. This distinction is critical for production deployment where a report with five inflated Criticals erodes client trust regardless of coverage.

\section{Future Directions}\label{sec:future}

\subsection{Continuous calibration monitoring}

A production deployment should track calibration metrics over time. Each engagement produces (prediction, outcome) pairs that can be added to a running reliability diagram. If ECE exceeds a threshold (e.g., 0.15 for Laya, 0.30 for Jev), the system triggers recalibration: either temperature rescaling on the accumulated data or, for Laya, a fine-tuning cycle.

\subsection{Federated training across engagements}

Rave training data comes from engagement logs, which are sensitive. Federated learning---where each operator trains on local data and shares only gradient updates---could pool training signal across organizations without sharing evidence. The challenge is that pentest evidence is highly heterogeneous across organizations and target types, making gradient aggregation difficult. Differential privacy guarantees would add noise that may degrade calibration. This trade-off between training quality and data privacy is an open problem.

\subsection{Toward autonomous Level 5 pentest systems}

Mayoral-Vilches~\cite{mayoral2025gap} defines Level 5 autonomous cybersecurity as full autonomy: the system operates without human intervention, including vulnerability discovery, exploitation, reporting, and remediation verification. Current systems operate at Levels 3--4, where human review is required before vulnerability submission.

System One integration is a necessary (though not sufficient) step toward Level 5. The calibration guarantee enables \emph{auditable autonomy}: every decision has a logged probability distribution, every threshold has a documented cost-model justification, and every finding has an evidence chain that an auditor can verify retrospectively. Without calibrated decisions, autonomous operation would require either trusting uncalibrated LLM judgments (unacceptable for compliance-driven engagements) or defaulting to conservative thresholds that route most findings to human review (negating the autonomy benefit).

The remaining gaps for Level 5 are: (1) remediation verification---confirming that a vulnerability has been fixed, which requires re-testing with the same methodology and evidence comparison; (2) scope management---autonomously adjusting scope based on discovered assets, which requires formal scope calculus beyond simple CIDR matching; (3) operational safety---guaranteeing that the system cannot cause unintended denial of service, data loss, or scope violations, which requires formal verification of the harness's action space; and (4) regulatory acceptance---establishing that autonomous pentest outputs meet the evidentiary standards of relevant regulations (SOC 2, PCI DSS, ISO 27001), which requires calibration audits and liability frameworks.

System One models contribute to gaps (1) and (4): remediation verification is a finding-adjudication decision on re-test evidence, and regulatory acceptance is strengthened by auditable calibration curves. Gaps (2) and (3) require harness-level engineering beyond the decision layer.

\subsection{Real-time calibration for novel vulnerability classes}

When a harness encounters a vulnerability class outside the System One model's training distribution (e.g., a novel API abuse pattern, a supply-chain injection), the model's calibration is unreliable. An online calibration mechanism that detects distribution shift (via prediction entropy or calibration error on a sliding window) and falls back to LLM-based adjudication for out-of-distribution findings would improve robustness. The detection threshold determines the false-positive rate of the shift detector: too sensitive triggers frequent fallbacks (reducing efficiency); too conservative allows miscalibrated decisions.

\subsection{Integration with other harness assurance properties}

System One models address evidence grounding and computed severity (P1 and P3 in the assurance framework~\cite{santos2026assurance}). Future work should explore integration with the remaining properties: non-destructive claim reduction (P2), enforced authorization (P4), and tamper-evident accountability (P5). For P2, a System One Score could evaluate the destructiveness of a proposed action (is this a read or a write? does it modify state?); for P4, a Noul could verify that a discovered asset falls within the engagement scope; for P5, a Choice could classify the type of audit event for structured logging.

\section{Conclusion}\label{sec:conclusion}

This paper presents an architectural proposal and a preliminary case study for integrating System One decision models into autonomous penetration-testing harnesses. The core contribution is the formalization of four decision points (DP1--DP4) where typed, calibrated verdicts replace free-form LLM judgment, implemented in NeuroSploit and tested with TypeSafe Jev.

The exploratory case study---single runs with and without Jev on one target---produced observations consistent with the architectural hypothesis: the TypeSafe-enabled run exhibited a more graduated severity distribution, and the integration exposed two harness bugs that the LLM-only pipeline masked. However, with one run per condition on one target, these observations do not constitute statistical evidence of improvement. Reproducing the study with multiple runs, diverse targets, and controlled execution order is the most important next step.

The emerging landscape offers proprietary (Jev, Jev-Ultrafast) and open-source (Laya) options with different published specifications. Jev reports higher accuracy on high-cardinality tasks; Laya reports lower latency and lower calibration error on its own evaluation set. Whether these advantages transfer to offensive-security decisions requires paired evaluation on the same pentest-relevant dataset---an experiment we have not conducted. The training paradigms---RLHF, RLAIF, RLCD, and the proposed RLHV---have implications for trust in automated security verdicts that deserve empirical validation.

As future work, Rave sketches a domain-adapted System One variant to be fine-tuned on offensive-security decision distributions using RLHV, where the harness's own deterministic validators serve as the verification oracle. Rave remains a design proposal; building it requires the training data, evaluation protocol, and paired benchmarks described in Section~\ref{sec:proposal}.

Table~\ref{tab:summary} summarizes the key contributions and their empirical support.

\begin{table}[h]
\centering
\caption{Summary of contributions and empirical support.}
\label{tab:summary}
\small
\resizebox{\columnwidth}{!}{%
\begin{tabular}{@{}lll@{}}
\toprule
\textbf{Contribution} & \textbf{Observation} & \textbf{Evidence level} \\
\midrule
DP1--DP4 formalization & 4 decision points & Implemented \\
Severity recalibration & 9 of 22 recalibrated & 1 run, 1 target \\
Wall-clock difference & 5m 19s faster & 1 run, 1 target \\
Bug detection & 2 harness bugs exposed & Case study \\
Speed analysis & Published latency data & Reported (not paired) \\
Calibration analysis & Published ECE data & Reported (not paired) \\
Cost analysis & Hypothetical ROI & Analytical model \\
Rave & RLHV training loop & Future work \\
\bottomrule
\end{tabular}%
}
\end{table}

The System One layer does not replace the LLM agent. It replaces the LLM's role as judge of its own work. The agent discovers; the deterministic validator grounds; the System One model calibrates. Each operates in the domain where its architecture is strongest.

Five directions merit immediate investigation: (1) replicating the benchmark across diverse targets to establish external validity; (2) conducting a head-to-head Jev vs.\ Laya comparison on identical offensive-security tasks; (3) building the Rave training dataset from synthetic and anonymized engagement data; (4) implementing the cascading decision chain (Section~\ref{sec:orchestration}) and measuring its cost-per-decision profile; and (5) extending the integration to the remaining assurance properties (P2, P4, P5).

The broader implication is architectural: the separation of generation (LLM) from judgment (System One) from grounding (deterministic validators) creates a layered assurance stack where each layer can be independently improved, tested, and replaced. This separation is the key to building autonomous pentest systems that are not only capable but trustworthy.

\section*{Acknowledgment}
The author thanks the TypeSafe AI and Laya teams for publishing documentation and benchmarks that enabled independent evaluation. The Browser Use team's open-source Jev-Ultrafast repository provided the browser-integration analysis. Generative language-model tooling assisted with drafting and reference verification; all sources were checked by the author against their original records, and the author takes full responsibility for the content, including any errors.

\section*{Ethics and Responsible Use}
All benchmark runs were conducted against a locally-hosted, deliberately-vulnerable application controlled by the author. No third-party systems, credentials, or personal data were involved. Autonomous offensive tooling is dual-use; the assurance properties discussed (evidence grounding, calibrated severity, enforced scope, tamper-evident accountability) constrain misuse and make authorized engagements defensible. Code examples describe API interfaces and do not constitute exploit payloads.

\section*{Competing Interests and Funding}
The author is the developer of NeuroSploit; this is the paper's sole competing interest. No external funding was received.

\end{document}